# An experiment in price perception error

**Shawn Berry, DBA[1*]**

[1]William Howard Taft University, Lakewood CO, USA

*Correspondence: shawnpberry@gmail.com

## Abstract

The process of consumer decision-making is multidimensional, and price perception is a very important but still not well-understood dimension for both marketers and consumers. Although heuristics or mental shortcuts are seen as biased and can cause decision errors, consumers tend to use price knowledge heuristics for purchase decisions, sometimes relying on old information. This study examined the effects of individual attitudinal and demographic factors on the respondents' ability to correctly guess the prices of 13 products and services (n=351) using ANOVA and the development of a seven-factor price perception model using latent variable analysis (lavaan). While most respondents preferred to either research prices first or compare prices to a similar product, they either systematically underestimated or overestimated the prices of the products. In the ANOVA, brand loyalty, importance of substitute products and knockoff products, how financially well-off the respondent household was growing up, product quality importance, haggling tendency, and level of income were statistically significant predictors of price perception error. Latent variable analysis revealed that demographics, decision making, and price sensitivity factors had the greatest influence on price perception error. Level of education and income, frequency of regret, coupon importance, and how respondents chose to save or spend money were significant latent variables. This study adds an important contribution to the body of knowledge on price perception, with directions for future research.

## 1. Introduction

### 1.1 Background and Rationale

Price perception by consumers remains a challenge to marketers and a source of inquiry for researchers, given that prices represent a primary means for consumers to make choices but are complicated by individual preferences and subjective factors. Price perception is the means by which consumers use and mentally process the prices of products and services (Zeithaml, 1988), but the discernment of which is affected by product quality attributes, mental price benchmarks, and tradeoffs of quality and price (Zong & Guo, 2022). Therefore, there has been continuous development in the modeling of price perception in the context of how price information is used and acted upon by consumers, and the ultimate effect upon how marketers and markets address these nuances (Monroe, 1973; Zeithami, 1988). This development of research into price perception has moved the school of thought from utility-based economic models (Monroe

& Petroshius, 1981), which have limitations to the realm of psychology that considers individual biases and qualitative preferences (Monroe & Krishnan, 1985; Thaler, 1985). Hence, price perception error is the disparity between the actual price of a product offering and how individual consumers regard prices as biased according to their preferences and cognition of price knowledge (Monroe & Krishnan, 1985; Monroe & Lee, 1999), specifically, the heuristic of mental processing, understanding, and recall of prices to make decisions (Dickson & Sawyer, 1990).

The implication of heuristics in the context of price perception errors is that consumers do not possess perfect price recall (Gabor and Grainger, 1961). Indeed, Tversky and Kahneman (1974) demonstrated that people tend to make errors when using heuristics for decision-making in an uncertain environment. These error tendencies also extend into the realm of business-to-business transactions, where purchasing agents at companies may also rely on heuristics when it comes to prices (Saab and Botelho, 2020). In fact, Dickson and Sawyer (1990) suggested that price knowledge among consumers is generally limited to the extent that many shoppers were unaware of the actual price they were going to pay for the products that were briefly selected. Price knowledge heuristics by consumers sometimes rely on knowing prior prices that consumers use as an internal reference (Helson, 1964; Monroe, 1973), wherein newly encountered prices are either thought of as being cheaper or more costly than they were used to (Niedrich et al. (2001)). As a result, consumers adapt to new information, and this becomes the new internal reference price, a process that has been validated by several scholars (Kalyanaram and Winer, 1995; Mazumdar et al., 2005). This can lead to price anchoring, wherein a given price can bias decisions (Zong and Guo, 2022). Another heuristic for price knowledge that leads to price perception errors deals with price range. Whereas consumers may not know a specific price point, it is possible for a consumer to subjectively misinterpret a price given that they may generally know the price range of a product offering (Volkmann, 1951; Niedrich et al., 2001). Therefore, if a consumer is comparison shopping, the price of a product offering is tempered not just by the maximum and minimum prices that the consumer has seen (Niedrich et al., 2001), but also by where that price falls in the overall ranking of prices that the consumer knows about (Parducci, 1965). On this basis, errors in price perception can occur because of biases arising from internal reference price cognitions. Furthermore, incorrect assumptions about prices can be triggered by the heuristic that high prices imply a high offering quality (Gneezy et al., 2014). Alternately, in a retail setting, consumers may make decisions according to a heuristic that relies upon their perception of how that retailer traditionally sets its prices in the context of whether prices are generally perceived as being low or not (Hamilton & Urminsky, 2013)

Models of price perception error fall into two categories: psychological models of individual price perception or reference prices, and mathematical models to predict market behavior (Mazumdar et al., 2005). Among the mathematical modeling approaches, structural equation modeling (SEM) has been successfully used to examine reference prices and price perception to determine the strength and validity of relationships among variables in a given conceptual framework, specifically among the observed variables (measurement model) and latent variables (structural model) (Lichtenstein et al., 1993; Niedrich et al., 2001). In particular, latent variable analysis (lavaan) has been successfully used in studies on price perception (Lin, 2015; Tu, 2022; Vellido, 2000; Victor, 2018; Wang, 2023; Yasri et al., 2020; Zhao et al., 2021). Meta-analysis also provides depth of understanding and validation of findings about price perception in the context of how consumers obtain information, branding, and product quality (Völckner and Hofmann, 2007), and product categories (Kalyanaram and Winer, 1995). Analysis of variance (ANOVA) has also been employed in price perception studies (Chen, Zhang, & Wu, 2024; Levrini & dos Santos, 2021; Lii, 2009; Ma, 2018; Shirai, 2015; Shirai, 2017).

A number of factors and variables influence price perception and, ultimately, consumer behavior. Logically, these multidimensional factors also affect price perception errors. First, demographic variables have a significant impact on price perception. Age affects price perception in that younger consumers are less accurate in price perceptions than older consumers, whereas the latter have acquired more knowledge of prices over time (Estelami & De Maeyer, 2004) and evaluate prices with respect to quality, and the former are affected by social media and fashion movements (Leivang & Sharma, 2023). Gender also plays a role in price perception, with women generally being more accurate than men with respect to price perception due to greater price comparison and placing a greater priority on price during shopping across various categories (Leivang & Sharma, 2023). However, men tend to have more price knowledge of their products (Leivang & Sharma, 2023). Consumers' level of education affects price perception in that those possessing higher levels of education made fewer errors and showed more price knowledge, possibly due to greater cognitive performance and financial savvy (Gabor & Grainger, 1961). The income level of consumers is also a determinant of price perception, since affluent consumers were not price-sensitive but viewed prices as a trade-off with quality and benefits to the extent of prioritizing value over price in some instances (Gabor & Grainger, 1961). Region of residence may also confound price perception among consumers, especially given variations in the cost of living and economic environment in different areas (Gabor & Grainger, 1961). Product quality has been shown to influence price perception, since price is often used as a proxy for quality, implying that higher prices must mean higher quality (Zeithaml, 1988). Whereas with intangible or experience purchases, quality is assessed after the purchase, and this greatly reinforces the linkage between price and quality as opposed to tangible goods, where quality can be assessed before purchasing (Zeithaml, 1988). However, the assessment of product quality and price perception is moderated according to product knowledge, where more experienced consumers will not rely on price as a proxy for quality as opposed to relatively inexperienced consumers (Zeithaml, 1988). The tendency of consumers to shop online has implications for price perception. Consumers who frequently shop online tend to have better price knowledge than consumers who do not frequently shop online (Fenneman et al., 2022). The country of origin of a product has been found to influence price perception when consumers have pre-existing notions that certain countries make inferior or superior goods according to the perceived quality associated with goods from those countries, and therefore, subsequently attach either a perceived premium or devaluation for goods from a given country (Diamantopoulos et al., 2021). Haymond (2022) suggests that the socioeconomic background of a consumer has a bearing on how any given consumer perceives prices, where those from less-well-off households tend to be more price-sensitive and unwilling to pay premium prices than those from well-off households who are more willing.

Second, individual decision-making attributes are factors in price perception. Consumers with higher levels of self-confidence in decision-making tend to use price anchoring less than consumers with lower levels of self-confidence, implying that individual self-confidence can reduce the bias in price perception (Zong & Guo, 2022). Consumers' levels of optimism toward their future have an impact on their decision-making behavior to the extent that those consumers who exhibit very high levels of optimism tend to not be cautious in their decisions as compared to those with medium levels of optimism who exhibit sensible decision-making (Puri & Robinson, 2007). Levels of money savvy among consumers affect price perception, since higher levels of money savvy imply greater accuracy with price data due to better financial proficiency and, therefore, better decision-making (Hashmi et al., 2021). The extent to which consumers feel financially secure impacts price perception, such that those consumers who see themselves as less financially secure will tend to be less happy, and therefore, more judgmental or hypercritical of prices, effectively changing their price perception (Dias et al., 2021). Coupons change consumer price perception contextually in terms of how quickly savings can be realized with a coupon compared to a rebate (Folkes & Wheat, 1995). The amount of the coupon discount effectively signals if the price is perceived as being high (Raghubir, 1998), and that large coupon amounts lead to greater scrutiny of the offer by consumers and a greater likelihood of purchase. Finally, brand-related and conspicuous consumption factors affect price perceptions. Brand loyalty, as expressed by a consumer's continued desire to buy products or services, implies that those consumers are less likely to care about premium prices, and therefore may ignore the possibility of price perception bias (García-Salirrosas et al., 2024). Consumers that engage in status seeking and signaling wealth or material possessions affect price perception. When those consumers view the price as a barrier to exclude others from being able to make a purchase, this results in conspicuous consumption given that the price is a signal that denotes the product is somehow exclusive, and the display of these possessions is a behavioral signal to others in society (Zhang, 2022). Therefore, this tendency to pay higher prices to achieve conspicuous consumption implies that the behavior overrules the tendency to be sensitive to high prices (Baysal Kurt & Kara, 2024). The purchase of counterfeit products or knockoffs has an effect on price perception because consumers frame the original branded product price as being overpriced compared to the lower price for the knockoff product on the assumption that they are of similar quality, even though the knockoff is of inferior quality (Ndofirepi et al., 2022). In other words, knockoffs disrupt the value proposition of owning the original brand name product by distorting the quality-price paradigm. The availability of substitute products also has an effect on price perception, wherein the consumer may regard a lower-priced substitute product of the same utility as being fairer than the brand name product, and therefore, reframe the price of the brand name product as being unreasonably priced (Yasri et al., 2020). As a result, price perception errors due to substitute products can create a problem for consumers, in that both relative and absolute price differences among substitutes themselves complicate consumer choices (Azar, 2011). Haggling behavior has the effect of reframing price perception because prices themselves are not seen as absolute but as a suggested point from which to negotiate, and therefore, where the buyer feels the final price is fair and satisfactory (Zhang & Jiang, 2014). Mukherjee et al. (2021) studied the role of consumer regret and found that consumers tended to change their purchase behavior and intentions when they perceived disparities in prices, resulting in greater regret when prices were perceived as being high. Chandrashekaran (2020) found that price perception was affected by the propensity of consumers to save or spend their money such that prices were viewed more positively by consumers that preferred to spend versus consumers that preferred to save, and suggested that the reluctance to spend implies an aversion to potential discomfort by those consumers. The role of limited-time offers on price perception was studied by Sugden, Wang, and Zizzo (2019), and they found that the overarching call to an urgent decision creates a psychological disparity between the actual needs of the consumer and overcoming the fear that the consumer will somehow lose value by missing out on a time-sensitive deal on something that was not actually more valuable. As a result, the price perception error from a limited-time offer occurs because the special price distorts the true value of the product as compared to the original price in the process (Sugden, Wang, & Zizzo, 2019). Finally, in a study by de Wulf et al. (2005), store brand products have been shown to be equal to or better than the quality of brand name products, and thus, are viable alternatives for consumers that are sensitive to prices given quality. Furthermore, this price-quality tradeoff of store brand versus brand name products changes the dynamics of price perception since the brand and store names moderate the willingness to pay higher prices since this may infer higher quality (Render & O'Connor, 1976).

In general, price perception can be described as an aspect of consumer behavior wherein consumers appraise the price of goods and services during their purchase journey using a variety of qualitative and quantitative criteria to make a decision. Quantitatively, this appraisal can take the form of actively looking up or searching for prices from a variety of information sources or through some mental shortcut that relies on the knowledge of the price of a recent purchase or similar product. Qualitatively, the price of an offering is judged by the consumer relative to products or services of similar quality or benefits, products or services of higher or lower quality that are adjacent to the product (e.g., a mid-level offering versus a high- or entry-level offering). The consumer interprets this information to make a judgment whether an offering is, in their mind, a "good deal," meaning that the consumer may or may not find it to be reasonably priced. In this study, the difference between the actual price of the offering and the price that the consumer has in mind according to their shopping heuristic or mental shortcut is referred to as the price perception error.

### 1.2 Research Problem and Objectives

This paper will examine the main research question, which is how price perception error is influenced by factors that shape consumer price perception. The objectives of this study are to identify statistically significant factors and variables that influence price perception error. The research question will also be examined in the context of what heuristic consumers use as a mental guidepost for price knowledge and which approaches yield the lowest price perception error. Using the conceptual framework presented herein, a seven-factor model is proposed, and both analysis of variance (ANOVA) and latent variable analysis (lavaan) are employed to determine the extent to which latent variables affect their respective factors, and which factors are the most influential with respect to the under- and overestimation of prices for a variety of products and services.

### 1.3 Significance and Contribution

This study will contribute to the body of research on price perception and provide a greater understanding of why consumers make errors through heuristic shortcuts that potentially and wrongly inform whether a price is a "good deal." The methodology used herein illustrates an approach to testing the extent of price perception error for different kinds of purchases by measuring a variety of consumer dimensions and codifying their price knowledge heuristics. These insights are valuable to marketers and consumer behavior analysts because the triggers for these errors are identified for products and services of varying complexity.

### 1.4 Paper Structure

The remainder of this paper is organized as follows. First, prior work by scholars in the area of price perception is discussed, focusing on the central themes and frameworks for price perception that have emerged and evolved over time. Second, the current study will be introduced, including the methodology and conceptual framework that was employed. Third, the results of the study are provided, followed by a discussion. Finally, the paper concludes with a brief discussion of its limitations and paths for future research into price perception error.

## 2. Materials and Methods

The data used in this study were collected using a panel recruited from the crowdsourcing platform Clickworker.com between February 21 and 26, 2025. 70 respondents aged 18 years and over from the United States agreed to participate. Some of the respondents were excluded from the study. Only one person did not agree to participate when asked to agree to the honor code of not using tools to assist with the survey and was not shown the survey. Of the 369 respondents who gave consent, one was excluded due to not being from the United States sample population (failed the regional check), five failed attention check questions, seven failed to enter the correct confirmation code at the end of the survey, and 12 were caught trying to take the survey more than once. The final sample size was 351.

The questionnaire, illustrated in Appendix A, used a Google Form with multiple choice and short text answer fields and consisted of 58 questions and one final field for the respondent to enter their confirmation code. Of the 58 questions, 13 required respondents to enter their best guess of a price from a range of products and services as an open-ended response, and the remainder were closed-ended questions consisting of 5-point Likert scale responses to measure attitudes and habits and multiple choice selections for the collection of categorical and demographic data. Multiple choice questions were used for attention check questions at various points in the survey, consisting of challenging the respondent to select the item that does not belong in the list of choices.

The price perception experiment questions consisted of 13 products and services presented from low-value and frequently purchased items to progressively more complex, less tangible, or less-frequently purchased but higher-valued items. The products, in order, were 1 lb of sweet organic onions, Vans shoes, 1 lb of Domino granulated sugar, dishwasher, electric dryer, 50-inch flatscreen television, couch, Nike Airforce shoes, Zegna men's boots, city bus fare, rate for one-night in an economy hotel, one-way economy airfare from New York JFK to Los Angeles LAX airport, and round-trip economy airfare from New York JFK to Paris CDG airport. Pictures of the items or illustrations of the services were shown to the respondents, with the exception of two branded products, Domino sugar and Nike shoes. Brand names were not used in the question, and branding was removed from the pictures to appear generic. Respondents were instructed to enter their best guess in US dollars (or dollars and cents, where applicable, for low-value items).

The actual prices of the products were collected online by sampling various retailers and airlines during the week of February 21, 2025. In the case of brand-specific products, the range of retailers offering those products was generally limited to the manufacturer or exclusive retailers, and therefore, the number of prices sampled was fewer. An example is Zegna, a luxury brand. The prices for each product or service were averaged. Airline prices were sampled from Skyscanner.com. Durable goods prices were sampled through the US versions of the websites for Costco, Lowes, Menards, Home Depot, Wal-Mart, and Target. Furniture prices were sampled through US websites for Abt, Best Buy, Bob's Discount Furniture, JC Penney, Lowes, Ikea, Target, Price Busters Furniture, and US Appliance. Transit fares in the United States were sourced from Picodi.com. Food prices were also sampled through Instacart stores and websites for regional US grocers and retail stores: Dollar Tree, Shaw's, Meijer, Publix, Albertsons, Piggly Wiggly, Harris Teeter, Acme Markets, Food Bazaar, Tony's Fresh Market, The Kosher Marketplace, Giant Eagle, Central Market, Butcher Shop

Direct, Rite Aid (before their bankruptcy declaration), Shopper's Food, Fred Meyer, Gerbes Super Markets, Boxed Greens, Ralph's, King Soopers, Safeway, Bristol Farms, and Sprouts Farmers Market. Shoe prices were sampled from Nike.com, Zegna.com, Journeys.com, ASOS.com, Hibbett.com, Modesens.com, Vans.com, Farfetch.com, Lyst.com, Soxy.com, and Yoox.com.

Table 1 presents the variables used in this study.

**Table 1.** Variable list

| Label | Description | Model Role | Variable Type |
|---|---|---|---|
| Price Perception Error | Actual price – Guessed price | Dependent Variable | Continuous |
| StatusBrands | Consumer preferences related to brand attributes | Latent Factor | N/A |
| - Wearing brand names | Importance of wearing or displaying brand names | Observed Variable | 5-point Likert |
| - Brand Loyalty | Degree of loyalty to brands | Observed Variable | 5-point Likert |
| - Trendy | Importance of being a trend follower | Observed Variable | 5-point Likert |
| - Showing off wealth | Importance of signaling material wealth | Observed Variable | 5-point Likert |
| DecisionMaking | Consumer decision processes | Latent Factor | N/A |
| - Decision Confidence | Self-assessed ability to make decisions | Observed Variable | 5-point Likert |
| - Optimism | Self-assessed sense of optimism toward life | Observed Variable | 5-point Likert |
| - Money Wise | Financial acumen in evaluating and choosing purchases | Observed Variable | 5-point Likert |
| - Regret Frequency | Frequency of regretting purchase decisions | Observed Variable | 5-point Likert |
| OnlineShopping | Digital purchasing behaviors | Latent Factor | N/A |
| - Online Frequency | How often consumers shop online | Observed Variable | 5-point Likert |
| - Availability online | Importance of products being available online | Observed Variable | 5-point Likert |
| PriceSensitivity | Responsiveness to price changes | Latent Factor | N/A |
| - Substitute Importance | Importance of substitute products to respondent | Observed Variable | 5-point Likert |
| - Knockoff Importance | Importance of knockoff products to respondent | Observed Variable | 5-point Likert |
| - Coupon Importance | Importance of coupons to respondent | Observed Variable | 5-point Likert |
| - Store Brands Superior | Attitude inferiority/superiority of store brand products | Observed Variable | 5-point Likert |
| QualityFocus | Quality-based evaluations | Latent Factor | N/A |
| - Quality Importance | Importance of quality over price | Observed Variable | 5-point Likert |
| - In-store Shopping Frequency | How often the respondent shopped in brick & mortar stores | Observed Variable | 5-point Likert |
| FinancialSecurity | Perception of personal financial comfort & stability | Latent Factor | N/A |
| - Employment Secure | Attitude toward respondent's own employment situation | Observed Variable | 5-point Likert |
| Demographics | Control variables representing basic consumer characteristics | Control Variables | N/A |
| - Age | Consumer's age | Observed Variable | Categorical |
| - Gender | Consumer's gender identity | Observed Variable | Categorical |
| - Income | Annual household income | Observed Variable | Categorical |
| - Education | Highest education level attained | Observed Variable | Categorical |
| - Region | Geographic region of residence in the USA | Observed Variable | Categorical |

The data were coded from the multiple-choice answer responses via the variable coding scheme displayed in Table 2.

**Table 2.** Variable coding scheme.

| Variable name | Variable description | Variable type | Coding |
|---|---|---|---|
| Age | Age group of respondent, years of age | Categorical | 18–24 = 1<br>25–34 = 2<br>35–44 = 3<br>45–54 = 4<br>55 and older = 5 |
| Gender | Gender of respondent | Categorical | Female = 0 |
| | | | Male = 1<br>Non-binary = 2 |
| Income | Annual income level of respondent, USD | Categorical | Less than $30,000 = 1<br>$30,000–$49,999 = 2<br>$50,000–$69,999 = 3<br>$70,000 and over = 4 |
| Education | Education level of respondent | Categorical | Did not finish high school = 0<br>High school graduate = 1<br>Some college = 2<br>Bachelor's degree = 3<br>Master's degree = 4<br>Postgraduate or higher = 5 |
| Region | United States region of residence of respondent | Categorical | Middle Atlantic = 1<br>New England = 2<br>South Atlantic = 3<br>East South Central = 4<br>West South Central = 5<br>Mountain = 6<br>Pacific = 7 |
| Likert scores | Degrees of agreement or importance of behavioral factors to measure respondent attitudes | Ordinal | Definitely unimportant/disagree = 1<br>Somewhat unimportant/disagree = 2 |

| | |
|---|---|
| | Neither agree nor disagree/important nor unimportant = 3<br>Somewhat important/agree = 4<br>Definitely important/agree = 5 |

**Adapted from Berry (2024)**

The seven-factor model of price perception error is shown in Figure 1.

**Figure 1.** Methodological framework

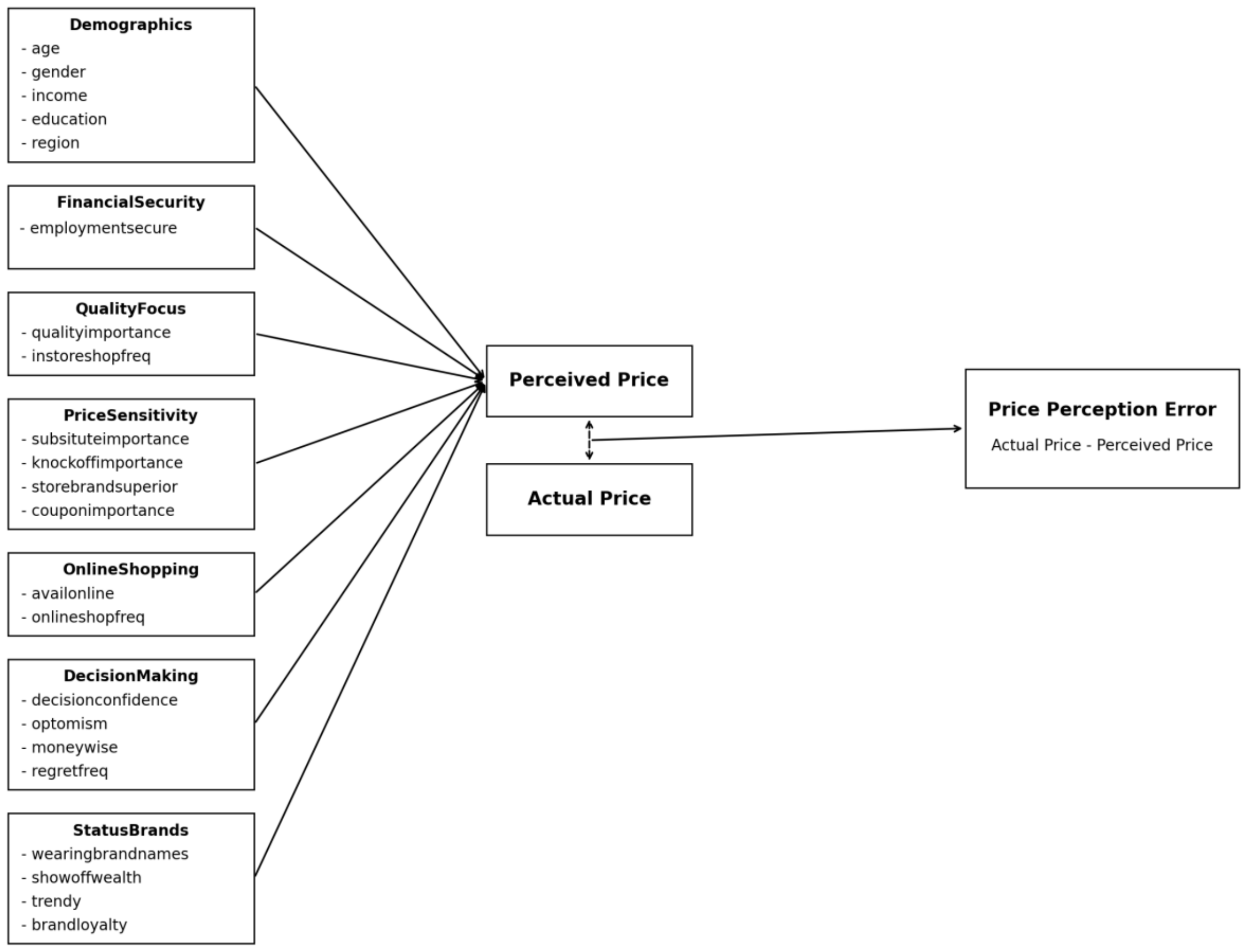

**Source:** data analysis

## 3. Results

The sample data were analyzed and modelled.

### *3.1. Descriptive statistics*

Table 3 displays the frequency distributions of the respondents' demographic attributes. Over 85% of the respondents were between the ages of 25 and 54, and over a third of respondents were between 35 and 44 years old. Over 70% of the respondents were female. Over 65% of respondents had some college level education, including a bachelor's degree. Almost 65% of respondents earned $49,000 per year or less with almost 44% of respondents earning less than $30,000 per year. Almost 58% of the respondents were from the South Atlantic, Middle Atlantic, and West South Central regions of the United States.

**Table 3.** Demographic attributes of the respondents

| Variable | Category | N | % of total |
|---|---|---|---|
| Age | 18-24 | 28 | 7.98 |
| | 25-34 | 86 | 24.50 |
| | 35-44 | 135 | 38.46 |
| | 45-54 | 78 | 22.22 |
| | 55 and older | 24 | 6.84 |
| Gender | Female | 248 | 70.66 |
| | Male | 102 | 29.06 |
| | Non-binary | 1 | 0.28 |
| Education | Some college | 141 | 40.17 |
| | Bachelor's degree | 89 | 25.36 |
| | High school graduate | 75 | 21.37 |
| | Master's degree | 30 | 8.55 |
| | Post-graduate degree | 10 | 2.85 |
| | Did not finish high school | 6 | 1.71 |
| Income | less than $30,000 per year | 153 | 43.59 |
| | $30,0000-$49,999 per year | 75 | 21.37 |
| | $70,000 per year or more | 75 | 21.37 |
| | $50,000-$69,999 per year | 48 | 13.68 |
| Region | South Atlantic (DE/DC/FL/GA/MD/NC/SC/VA/WV) | 76 | 21.65 |
| | Middle Atlantic (NY/NJ/PA) | 65 | 18.52 |
| | West South Central: (AR/LA/OK/TX) | 62 | 17.66 |

| | | |
|---|---|---|
| East South Central: (AL/KY/MS/TN) | 53 | 15.10 |
| Pacific: (AK/CA/HI/OR/WA) | 47 | 13.39 |
| Mountain: (AZ/CO/ID/MT/NV/NM/UT/WY) | 36 | 10.26 |
| New England: (CT/ME/MA/NH/RI/VT) | 12 | 3.42 |

**Source: data analysis**

Table 4 summarizes the descriptive statistics for the attitudinal variables, measured using a 5-point Likert scale. The five attitudes with the highest importance for respondents in descending order were saving money, price importance, quality importance, coupon importance, and online shopping frequency. The five least important attitudes for respondents in ascending order were showing off wealth, how much money was saved or spent when it was received (savespend), being trendy, status-seeking, and wearing brand names. The only variable that was not a 5-point Likert scale was savespend, an indicator that had three categories of how respondents chose to either save or spend all of their money, with the middle value being saving some money and spending the rest.

**Table 4.** Descriptive statistics – Likert scores

| Variable | n | Mean | SD | Median | Min | Max | SE |
|---|---|---|---|---|---|---|---|
| saving money | 351 | 4.73 | 0.73 | 5.00 | 1.00 | 5.00 | 0.04 |
| moneywise | 351 | 3.65 | 1.02 | 4.00 | 2.00 | 5.00 | 0.05 |
| brandloyalty | 351 | 3.63 | 0.95 | 4.00 | 1.00 | 5.00 | 0.05 |
| decisiontiming | 351 | 2.93 | 1.09 | 3.00 | 1.00 | 5.00 | 0.06 |
| subsituteimportance | 351 | 3.51 | 1.08 | 4.00 | 1.00 | 5.00 | 0.06 |
| availonline | 351 | 3.87 | 1.03 | 4.00 | 1.00 | 5.00 | 0.05 |
| knockoffimportance | 351 | 3.05 | 1.09 | 3.00 | 1.00 | 5.00 | 0.06 |
| optimism | 351 | 3.67 | 1.22 | 4.00 | 1.00 | 5.00 | 0.06 |
| onlineshopfreq | 351 | 4.12 | 0.95 | 4.00 | 1.00 | 5.00 | 0.05 |
| storebrandsuperior | 351 | 2.92 | 0.70 | 3.00 | 1.00 | 5.00 | 0.04 |
| limitedtimeoffer | 351 | 3.51 | 1.12 | 4.00 | 1.00 | 5.00 | 0.06 |
| origincountry | 351 | 3.02 | 1.21 | 3.00 | 1.00 | 5.00 | 0.06 |
| parentswelloff | 351 | 2.87 | 1.11 | 3.00 | 1.00 | 5.00 | 0.06 |
| qualityimportance | 351 | 4.56 | 0.67 | 5.00 | 2.00 | 5.00 | 0.04 |
| buyknockoff | 351 | 3.42 | 1.33 | 4.00 | 1.00 | 5.00 | 0.07 |
| instoreshopfreq | 351 | 3.61 | 1.25 | 4.00 | 1.00 | 5.00 | 0.07 |
| priceimportance | 351 | 4.63 | 0.60 | 5.00 | 2.00 | 5.00 | 0.03 |
| statusseeking | 351 | 2.41 | 1.29 | 2.00 | 1.00 | 5.00 | 0.07 |
| wearingbrandnames | 351 | 2.48 | 1.26 | 2.00 | 1.00 | 5.00 | 0.07 |
| showoffwealth | 351 | 1.89 | 1.11 | 1.00 | 1.00 | 5.00 | 0.06 |
| savespend | 351 | 2.08 | 0.48 | 2.00 | 1.00 | 3.00 | 0.03 |
| haggler | 351 | 2.58 | 1.30 | 2.00 | 1.00 | 5.00 | 0.07 |
| trendy | 351 | 2.16 | 1.16 | 2.00 | 1.00 | 5.00 | 0.06 |
| decisionconfidence | 351 | 3.70 | 0.99 | 4.00 | 1.00 | 5.00 | 0.05 |
| couponimportance | 351 | 4.17 | 0.97 | 4.00 | 1.00 | 5.00 | 0.05 |
| priceknowledge | 351 | 3.52 | 1.27 | 3.00 | 1.00 | 5.00 | 0.07 |
| riskaversion | 351 | 3.30 | 1.09 | 4.00 | 1.00 | 5.00 | 0.06 |
| regretfreq | 351 | 2.71 | 1.08 | 3.00 | 1.00 | 5.00 | 0.06 |
| splurge | 351 | 2.76 | 1.17 | 2.00 | 1.00 | 5.00 | 0.06 |
| employmentsecure | 351 | 2.94 | 1.38 | 3.00 | 1.00 | 5.00 | 0.07 |

**Source: data analysis**

Table 5 displays the guessed versus mean actual prices by respondents for various products and services and related descriptive statistics. The products and services with the lowest variability in guessed prices were onions, sugar, and a city bus ride in ascending order. The three products and services with the greatest variability were, in descending order, round trips from JFK to CDJ Paris, dishwasher, and dryer. The lowest standard errors of price guesses were for onions, sugar, and bus rides. The greatest standard error of price guesses were for round-trip economy airfare from New York to Paris, dishwashers, and couches.

**Table 5.** Price guesses by respondents

| Product/service | Mean actual | SD actual | Mean guess | Median guess | SD guess | SE guess | Min guess | Max guess | Shapiro-Wilk | p-value |
|---|---|---|---|---|---|---|---|---|---|---|
| 1 lb organic sweet onions | 1.90 | 0.30 | 2.65 | 2.49 | 1.53 | 0.08 | 0.15 | 9.89 | 0.931 | <.001 (***) |
| Vans shoe | 74.57 | 23.09 | 36.45 | 30.00 | 19.84 | 1.06 | 10 | 150 | 0.912 | <.001 (***) |
| 1 lb Domino granulated sugar | 2.34 | 0.43 | 3.27 | 3.00 | 1.88 | 0.10 | 1 | 14 | 0.818 | <.001 (***) |
| Dishwasher | 705.76 | 335.10 | 538.56 | 400.00 | 410.51 | 21.91 | 100 | 3000 | 0.736 | <.001 (***) |
| Dryer | 580.94 | 149.54 | 497.84 | 400.00 | 342.93 | 18.30 | 100 | 3000 | 0.772 | <.001 (***) |
| 50 inch flatscreen tv | 414.03 | 304.85 | 374.05 | 300.00 | 238.46 | 12.73 | 100 | 2000 | 0.764 | <.001 (***) |
| Couch | 655.74 | 426.53 | 507.75 | 399.99 | 399.75 | 21.34 | 100 | 3000 | 0.748 | <.001 (***) |
| Nike Airforce Shoe | 131.00 | 11.94 | 79.97 | 79.00 | 37.51 | 2.00 | 15 | 350 | 0.855 | <.001 (***) |
| Zegna boots | 1399.20 | 613.99 | 90.91 | 60.00 | 92.73 | 4.95 | 25 | 1000 | 0.535 | <.001 (***) |
| City bus ride | 1.77 | 0.62 | 2.70 | 2.00 | 1.81 | 0.10 | 1 | 10 | 0.737 | <.001 (***) |
| One night economy hotel stay | 158.45 | na | 96.87 | 85.00 | 49.95 | 2.67 | 25 | 500 | 0.827 | <.001 (***) |
| One-way economy JFK-LAX | 579.13 | 234.13 | 333.52 | 300.00 | 212.03 | 11.32 | 50 | 1500 | 0.854 | <.001 (***) |
| Round-trip economy JFK-CDG | 730.33 | 129.96 | 912.58 | 700.00 | 674.39 | 36.00 | 200 | 5800 | 0.776 | <.001 (***) |

**Source: data analysis**

Figure 2 illustrates a plot of the mean actual and guess prices for each product and service. Data points that were close to or on the perfect guess line suggest that respondents were accurate or close to their guesses. Points further away from the perfect guess line suggest that the respondents were not accurate. The plot illustrates that respondents were generally accurate with price guesses for commoditized products such as television, onions, sugar, and non-luxury footwear, but were very inaccurate when it came to trips and luxury footwear. Interestingly, guesses for household durable goods such as dryers, couches, and dishwashers were reasonably close.

**Figure 2.** Plot of mean actual price versus mean guess price

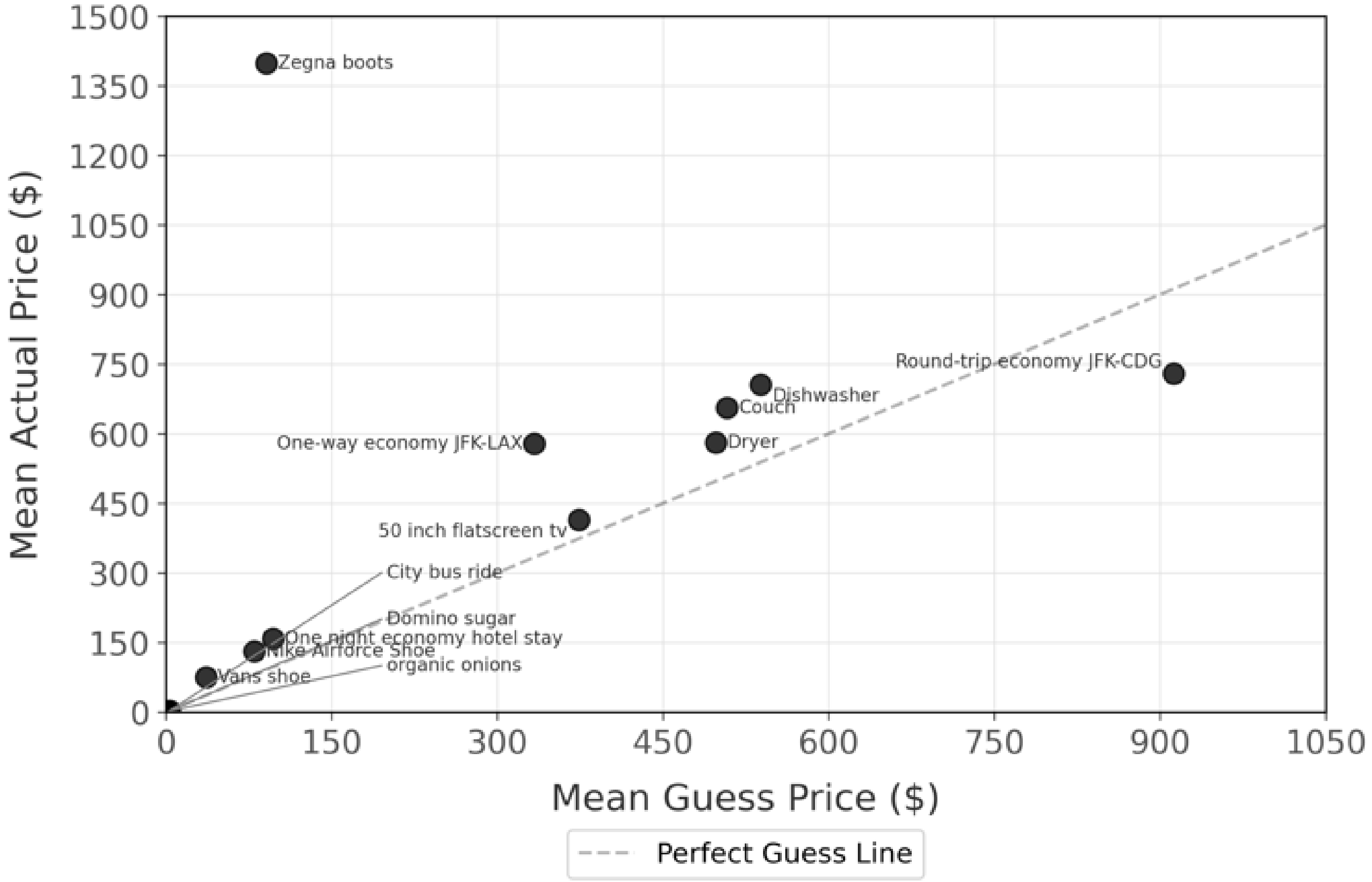

**Source: data analysis**

Table 6 displays the descriptive statistics for respondents' price perception errors. The price perception error is computed as

Price perception error = (actual price – price guess)/actual price.

Positive values indicate underestimated prices, and negative values indicate overestimated prices. The mean and median errors suggest that respondents overestimated prices for onions, Domino sugar, city bus fares, 50-inch televisions, and round-trip economy airfare from New York JFK to Paris. Respondents greatly underestimated the prices of Zegna boots (a luxury brand) and generally for the shoe brands Nike and Vans. For durable household goods, respondents showed modest underestimation of prices for dryers, dishwashers, and couches. Respondents also underestimated the prices of economy hotel stays and domestic US flight fares.

**Table 6.** Descriptive statistics of price perception errors by respondents

| Variable | n | Mean | SD | Median | Min | Max | SE | Shapiro-Wilk | p-value |
|---|---|---|---|---|---|---|---|---|---|
| onion error | 351 | -0.40 | 0.81 | -0.31 | -4.21 | 0.92 | 0.04 | 0.931 | <.001 (***) |
| vans error | 351 | 0.51 | 0.27 | 0.60 | -1.01 | 0.87 | 0.01 | 0.912 | <.001 (***) |
| 1 lb Domino error | 351 | -0.40 | 0.80 | -0.28 | -4.98 | 0.57 | 0.04 | 0.818 | <.001 (***) |
| Dishwasher error | 351 | 0.24 | 0.58 | 0.43 | -3.25 | 0.86 | 0.03 | 0.736 | <.001 (***) |
| Dryer error | 351 | 0.14 | 0.59 | 0.31 | -4.16 | 0.83 | 0.03 | 0.772 | <.001 (***) |
| 50-inch tv error | 351 | -0.20 | 0.83 | 0.03 | -6.25 | 0.76 | 0.04 | 0.764 | <.001 (***) |
| Couch error | 351 | 0.23 | 0.61 | 0.39 | -3.57 | 0.85 | 0.03 | 0.748 | <.001 (***) |
| Nike error | 351 | 0.39 | 0.29 | 0.40 | -1.67 | 0.89 | 0.02 | 0.855 | <.001 (***) |
| Zegna error | 351 | 0.94 | 0.07 | 0.96 | 0.29 | 0.98 | 0.00 | 0.535 | <.001 (***) |
| City bus error | 351 | -0.52 | 1.02 | -0.13 | -4.65 | 0.44 | 0.05 | 0.737 | <.001 (***) |
| Economy hotel error | 351 | 0.39 | 0.32 | 0.46 | -2.16 | 0.84 | 0.02 | 0.827 | <.001 (***) |
| JFK-LAX error | 351 | 0.42 | 0.37 | 0.48 | -1.59 | 0.91 | 0.02 | 0.854 | <.001 (***) |
| JFK-CDG error | 351 | -0.25 | 0.92 | 0.04 | -6.94 | 0.73 | 0.05 | 0.776 | <.001 (***) |

**Source: data analysis**

***Mean price guess error according to respondent attribute levels***

The mean price guess errors of respondents were evaluated with respect to their attributes and demographic characteristics using an analysis of variance (ANOVA) and Kruskal-Wallis (KW) tests. Table 7 displays the mean price guess error of respondents according to their Likert score level (1 through 5), with the exception of savespend, which was a 3-point Likert score that measured whether respondents saved all, some, or none of what they earned. Since the demographic factors were also coded according to the variable coding scheme in Table 2, Level 0 refers to the gender code for female respondents and Level 1 for male respondents.

The analysis results illustrate some interesting and statistically significant patterns that may help predict price perception error cues. First, brand loyalty was significant for both ANOVA and KW, showing that respondents tended to have smaller underestimation errors with increasing levels of brand loyalty. Second, the importance of substitute products for respondents was significant in the ANOVA and KW tests. With the exception of those in Level 2, which overestimated prices (somewhat unimportant), respondents generally had larger mean underestimation errors with increasing substitute importance levels. Third, the importance of knockoff products was statistically significant in the ANOVA and KW tests, illustrating an increasing underestimation of prices with increasing levels of importance of knockoffs. However, the exception was for respondents expressing that knockoffs were somewhat important to them (Level 4), as they had the greatest mean underestimation of prices. Fourth, the role of how well off the respondent's childhood household influenced mean price guess errors (parentswelloff) was statistically significant in the KW test. Respondents from impoverished households to some extent (definitely to somewhat impoverished, Levels 1 and 2) had greater mean underestimation errors than those from well-off households (somewhat to definitely well-off, Levels 4 and 5), with those from definitely well-off households having the lowest mean underestimation error. Fifth, the importance of product quality to respondents (qualityimportance) was statistically significant with the KW test, with respondents that expressed indifference (neither important nor unimportant, Level 3) having the largest mean underestimation error. The respondents from Level 2 (somewhat unimportant) had the lowest mean underestimation errors. Sixth, the tendency of respondents to haggle in purchase transactions (haggler) was statistically significant in both tests. Respondents that reported they did not haggle (Level 1) had the lowest mean underestimation error and those that reported frequent haggling (Level 5) had the highest mean underestimation error. Finally, income levels were statistically significant in the KW test, and the mean underestimation error decreased with increasing income levels.

Overall, the results suggest that respondents tend to underestimate prices, and despite some variables not being statistically significant, some key patterns are interesting. Respondents that were slow at timing their decisions (decisiontiming) had lower mean underestimation errors (Levels 1 and 2) than those that made fast decisions (Levels 4 and 5) who had much higher mean errors. Those respondents who saw themselves as definitely wise with money (Level 5 of moneywise) had lower mean underestimation errors than the other levels, with those who saw themselves as neither wise nor unwise having the greatest error (Level 3). The importance of wearing brand names for respondents showed curious results, with those placing great importance (Level 5) having the lowest mean underestimation error and the highest mean error at Level 4 (somewhat important). Respondents that saw themselves as definitely unwilling to take risks (Level 1 of risk aversion) had the greatest mean underestimation error as compared to those definitely willing to take risks (Level 5) who had the lowest mean error. Respondents with low levels of security (Levels 1 and 2) in their employment situation (employmentsecure) had higher mean underestimation errors than those with higher levels of security (Levels 4 and 5). Female respondents tended to have higher mean underestimation errors than male and non-binary respondents. Respondents who expressed higher levels of importance (Levels 4 and 5) for coupons (couponimportance) had greater mean underestimation errors than did those who expressed lower levels of coupon importance (Levels 1 and 2). Respondents at Level 5 (definitely important) for outward displays of wealth (showoffwealth) and trendy had the greatest mean underestimation error than those at Level 1 (definitely unimportant).

**Table 7.** Mean price guess error by Likert score level

| | | | | | Mean error | | | | | |
|---|---|---|---|---|---|---|---|---|---|---|
| **Variable** | **ANOVA F-value** | **ANOVA p-value** | **KW H-value** | **KW p-value** | **Level 0** | **Level 1** | **Level 2** | **Level 3** | **Level 4** | **Level 5** |
| moneywise | 0.361 | 0.781 | 3.782 | 0.286 | na | na | 0.113 | 0.132 | 0.122 | 0.085 |
| brandloyalty | 2.605 | 0.036* | 16.525 | 0.002** | na | 0.281 | 0.1 | 0.162 | 0.099 | 0.048 |
| decisiontiming | 0.627 | 0.644 | 1.507 | 0.825 | na | 0.062 | 0.099 | 0.113 | 0.148 | 0.114 |
| subsituteimportance | 2.83 | 0.025* | 10.395 | 0.034* | na | 0.144 | -0.008 | 0.11 | 0.132 | 0.161 |
| availonline | 0.27 | 0.897 | 3.874 | 0.423 | na | 0.139 | 0.062 | 0.124 | 0.11 | 0.122 |
| knockoffimportance | 3.057 | 0.017* | 10.819 | 0.029* | na | 0.045 | 0.069 | 0.099 | 0.184 | 0.071 |
| optimism | 1.085 | 0.364 | 3.888 | 0.421 | na | 0.136 | 0.067 | 0.072 | 0.12 | 0.148 |
| onlineshopfreq | 0.452 | 0.771 | 4.997 | 0.288 | na | -0.035 | 0.138 | 0.1 | 0.105 | 0.126 |
| storebrandsuperior | 0.27 | 0.897 | 0.96 | 0.916 | na | 0.168 | 0.107 | 0.11 | 0.124 | 0.206 |
| limitedtimeoffer | 1.034 | 0.389 | 4.53 | 0.339 | na | 0.21 | 0.078 | 0.09 | 0.114 | 0.135 |
| origincountry | 0.745 | 0.562 | 2.401 | 0.662 | na | 0.144 | 0.142 | 0.119 | 0.08 | 0.104 |
| parentswelloff | 2.325 | 0.056 | 10.37 | 0.035* | na | 0.065 | 0.175 | 0.092 | 0.117 | 0.022 |
| qualityimportance | 1.685 | 0.17 | 9.041 | 0.029* | na | na | 0.032 | 0.252 | 0.084 | 0.124 |
| buyknockoff | 4.213 | 0.002** | 12.881 | 0.012* | na | 0.053 | -0.003 | 0.134 | 0.119 | 0.191 |
| instoreshopfreq | 0.326 | 0.86 | 1.27 | 0.866 | na | 0.108 | 0.113 | 0.126 | 0.094 | 0.136 |
| priceimportance | 0.605 | 0.612 | 1.099 | 0.777 | na | na | -0.017 | 0.043 | 0.129 | 0.114 |
| statusseeking | 0.579 | 0.678 | 1.728 | 0.786 | na | 0.102 | 0.152 | 0.089 | 0.106 | 0.113 |
| wearingbrandnames | 1.874 | 0.115 | 5.398 | 0.249 | na | 0.132 | 0.118 | 0.053 | 0.153 | 0.002 |
| showoffwealth | 0.13 | 0.971 | 0.481 | 0.975 | na | 0.106 | 0.123 | 0.121 | 0.114 | 0.175 |
| savespend | 0.892 | 0.411 | 2.506 | 0.286 | na | 0.053 | 0.115 | 0.14 | na | na |
| haggler | 2.523 | 0.041* | 9.534 | 0.049* | na | 0.064 | 0.175 | 0.097 | 0.088 | 0.184 |
| trendy | 0.335 | 0.854 | 2.603 | 0.626 | na | 0.102 | 0.136 | 0.111 | 0.097 | 0.171 |
| decisionconfidence | 0.437 | 0.782 | 2.303 | 0.68 | na | 0.067 | 0.128 | 0.102 | 0.127 | 0.081 |
| couponimportance | 1.278 | 0.278 | 5.235 | 0.264 | na | 0.083 | 0.059 | 0.032 | 0.112 | 0.141 |
| priceknowledge | 1.116 | 0.349 | 4.291 | 0.368 | na | 0.117 | 0.058 | 0.152 | 0.102 | 0.096 |
| riskaversion | 0.402 | 0.807 | 3.3 | 0.509 | na | 0.158 | 0.113 | 0.145 | 0.105 | 0.085 |
| regretfreq | 1.021 | 0.397 | 4.747 | 0.314 | na | 0.109 | 0.133 | 0.104 | 0.123 | 0.001 |
| splurge | 1.264 | 0.284 | 4.973 | 0.29 | na | 0.125 | 0.085 | 0.148 | 0.144 | 0.037 |
| employmentsecure | 1.281 | 0.277 | 2.586 | 0.629 | na | 0.137 | 0.164 | 0.102 | 0.084 | 0.074 |
| age | 2.278 | 0.061 | 6.491 | 0.165 | na | 0.024 | 0.104 | 0.152 | 0.07 | 0.189 |
| education | 1.833 | 0.106 | 10.147 | 0.071 | na | 0.166 | 0.136 | 0.053 | 0.062 | 0.144 |
| income | 1.93 | 0.124 | 8.117 | 0.044* | na | 0.144 | 0.132 | 0.079 | 0.058 | na |
| region | 1.564 | 0.157 | 10.287 | 0.113 | na | 0.044 | 0.034 | 0.158 | 0.149 | 0.146 |
| gender | 1.242 | 0.29 | 3.656 | 0.161 | 0.129 | 0.077 | 0.071 | na | na | na |

**Source: data analysis**

*: $p<.05$. **: $p<.01$. ***: $p<.001$\

### *Error by price knowledge heuristic*

Table 8 displays the frequency distribution of the price knowledge heuristics used by respondents to mentally ascertain whether they think a price is a good deal. Most respondents declared that they researched prices first, followed by comparing a price to the prices of other similar products. Less than 8% of the respondents used the features or benefits of a product as the basis for knowing if they thought a price was a good deal for them. Just over 11% of respondents referred to their knowledge of the prices of Costco or Walmart, and about 11% of respondents used the recall of a price from a recent purchase of the same item.

**Table 8.** Price knowledge heuristic used by respondents

| Price knowledge heuristic | N | % |
|---|---|---|
| Look at features or benefits of product | 27 | 7.69% |
| Compare to Walmart or Costco prices for the exact same item/brand | 40 | 11.40% |
| Research prices first | 127 | 36.18% |
| Remember price from most recent purchase for the same item | 38 | 10.83% |
| Compare price to other products that are like the item I want | 119 | 33.90% |
| Totals | 351 | 100.00% |

**Source: data analysis**

Table 9 displays the percentage of respondents who used a given price knowledge heuristic according to their demographic characteristics. Among the heuristics, researching prices first and comparing a price to that of a similar product are the most popular approaches across demographic categories, whereas looking at the benefits or features of a product is the consistently least popular heuristic approach for respondents.

When price knowledge heuristics were examined with respect to age category, respondents aged 35 and over tended to research prices first, with 50% of those aged 55 and over having the greatest tendency. More than 40% of respondents aged 18 to 44 tended to compare the price to the price of similar products that they wanted. Notably, over 21% of respondents aged 18 to 24 tended to use the Costco and Walmart prices for comparison. The chi-square test of independence of age category versus price knowledge heuristic was not statistically significant, $X^2$ (16, $N$ = 351) = 22.296, $p$ = .136.

There are interesting patterns in the use of price knowledge heuristics with respect to gender. Among the heuristic categories, researching prices first was the most popular, with males preferring this approach (39.22%) over females (35.08%). The second most popular price knowledge heuristic was comparing the prices of similar products, with females having a greater preference (34.68%) than males (32.35%). All non-binary respondents preferred to recall the price from a prior purchase as a heuristic. The chi-square test of independence of gender versus price knowledge heuristic was not statistically significant, $X^2$ (8, $N$ = 351) = 11.79, $p$ = .160.

Respondents with greater levels of education generally tended to research prices first, with those at the post-graduate level being the most likely to do so (50%) and those that did not finish high school being the least likely (0%). Among those respondents that did not finish high school, their preferred price knowledge heuristics were comparing to prices of similar products (50%) and comparing to Walmart and Costco prices (33.3%), the greatest proportions in each of these categories. Interestingly, respondents with a master's degree had equal preference for either researching prices or comparing them to the price of a similar product (40% each). The chi-square test of the independence of the level of education of the respondents versus the price knowledge heuristic was not statistically significant, $X^2$ (20, $N$ = 351) = 20.99, $p$ = .398.

Respondents earning $70,000 per year or more tended to research prices the most (41.33%). The highest percentage of respondents relying on the recall of the price from a prior purchase was among those earning $50,000 to $69,999 per year (18.75%). Comparing prices to those of similar products appears to be the most preferred price knowledge heuristic among respondents earning $30,000 per year or less. However, both researching prices first and comparing prices of similar products seem to be equally preferred among respondents earning $30,000 to $49,999 per year. The chi-square test of the independence of the income level of the respondents versus the price knowledge heuristic was not statistically significant, $X^2$ (12, $N$ = 351) = 9.36, $p$ = .672.

Regional variations in how price knowledge heuristics are employed are observed. Those respondents that tended to research prices first the most were from the West South Central, South Atlantic, and New England states. The greatest proportion of respondents who tended to compare the prices of similar products were observed in the Middle Atlantic, followed by the East South Central and Mountain states. Researching prices first was also very popular among respondents from the West South Central, New England, South Atlantic, and Pacific regions. The chi-square test of the independence of the region of residence of the respondents versus the price knowledge heuristic was not statistically significant, $X^2$ (24, $N$ = 351) = 24.40, $p$ = .439.

**Table 9.** Percentage of respondents using price knowledge heuristics according to demographic characteristics

| Variable | Category | I look at the features or benefits of the product | I compare it to what the prices are at Walmart or Costco for the exact same item/brand | I research the prices first to get an idea | I remember the price from most recent purchase I made for the same item | I compare the price to other products that are like the item I want |
|---|---|---|---|---|---|---|
| Age | 18-24 | 7.14 | 21.43 | 21.43 | 7.14 | 42.86 |
| | 25-34 | 11.63 | 12.79 | 25.58 | 9.30 | 40.70 |
| | 35-44 | 8.89 | 11.11 | 41.48 | 11.11 | 27.41 |
| | 45-54 | 2.56 | 7.69 | 39.74 | 11.54 | 38.46 |
| | 55 and older | 4.17 | 8.33 | 50.00 | 16.67 | 20.83 |
| Gender | Female | 8.06 | 12.90 | 35.08 | 9.27 | 34.68 |
| | Male | 6.86 | 7.84 | 39.22 | 13.73 | 32.35 |
| | Non-binary | 0 | 0 | 0 | 100.00 | 0 |
| Education | Did not finish high school | 0 | 33.30 | 0 | 16.70 | 50.00 |
| | High school graduate | 5.30 | 9.30 | 38.70 | 6.70 | 40.00 |
| | Some college | 8.50 | 12.10 | 36.20 | 11.30 | 31.90 |
| | Bachelor's degree | 11.20 | 14.60 | 33.70 | 12.40 | 28.10 |
| | Master's degree | 0 | 3.30 | 40.00 | 16.70 | 40.00 |
| | Post-graduate degree | 10.00 | 0 | 50.00 | 0 | 40.00 |
| Income | less than $30,000 per year | 7.19 | 13.73 | 35.29 | 7.19 | 36.60 |
| | $30,0000-$49,999 per year | 9.33 | 9.33 | 33.33 | 14.67 | 33.33 |
| | $50,000-$69,999 per year | 8.33 | 10.42 | 35.42 | 18.75 | 27.08 |
| | $70,000 per year or more | 6.67 | 9.33 | 41.33 | 9.33 | 33.33 |
| Region | South Atlantic (DE/DC/FL/GA/MD/NC/SC/VA/WV) | 7.69 | 10.77 | 40.00 | 12.31 | 29.23 |
| | Middle Atlantic (NY/NJ/PA) | 8.33 | 8.33 | 33.33 | 0 | 50.00 |
| | West South Central: (AR/LA/OK/TX) | 10.53 | 5.26 | 43.42 | 11.84 | 28.95 |
| | East South Central: (AL/KY/MS/TN) | 11.32 | 16.98 | 18.87 | 9.43 | 43.40 |
| | Pacific: (AK/CA/HI/OR/WA) | 6.45 | 9.68 | 37.10 | 11.29 | 35.48 |
| | Mountain: (AZ/CO/ID/MT/NV/NM/UT/WY) | 5.56 | 16.67 | 30.56 | 5.56 | 41.67 |
| | New England: (CT/ME/MA/NH/RI/VT) | 2.13 | 14.89 | 42.55 | 14.89 | 25.53 |

**Source: Data analysis**

Table 10 displays the mean price perception error by respondents according to their stated method of price knowledge when it comes to knowing whether, in their mind, a price is a good deal. The largest mean overestimation errors (negative values) were observed among those respondents that tended to favor the Costco/Walmart price heuristic, particularly for city bus fares, Domino sugar, and round-trip airfares from JFK to Paris. The largest mean underestimation errors (positive values) were consistently observed for Zegna boots, a luxury product, across each price knowledge heuristic and, interestingly, the greatest value being observed among those respondents that favored the recall of prices of a recent purchase for the same item. The lowest overestimation errors were observed for those respondents who researched prices first in the 50-inch television and JFK to Paris flight category, and for respondents who recalled the price for a recent purchase of the same item for the JFK to Paris category. The lowest underestimation errors were generally observed for the dryer, couch, and dishwasher categories, mainly among respondents who favored the Walmart/Costco heuristic.

**Table 10.** Mean price perception error by price knowledge of respondents

| Mean error | Look at features or benefits of product | Compare to Walmart or Costco prices for the exact same item/brand | Research prices first | Remember price from most recent purchase for the same item | Compare price to other products that are like the item I want |
|---|---|---|---|---|---|
| Onion error | -0.350 | -0.566 | -0.373 | -0.325 | -0.400 |
| Vans error | 0.506 | 0.546 | 0.525 | 0.483 | 0.495 |
| 1 lb Domino error | -0.407 | -0.527 | -0.344 | -0.309 | -0.441 |
| Dishwasher error | 0.258 | 0.316 | 0.241 | 0.174 | 0.221 |
| Dryer error | 0.098 | 0.119 | 0.189 | -0.036 | 0.170 |
| 50-inch TV error | -0.265 | -0.237 | -0.138 | -0.453 | -0.165 |
| Couch error | 0.371 | 0.118 | 0.233 | 0.215 | 0.225 |
| Nike error | 0.424 | 0.409 | 0.403 | 0.362 | 0.370 |
| Zegna error | 0.926 | 0.932 | 0.937 | 0.941 | 0.934 |
| City Bus error | -0.666 | -0.814 | -0.459 | -0.505 | -0.467 |
| Economy hotel error | 0.407 | 0.460 | 0.393 | 0.360 | 0.365 |
| JFK-LAX error | 0.408 | 0.405 | 0.470 | 0.488 | 0.365 |
| JFK-CDG error | -0.183 | -0.405 | -0.101 | -0.075 | -0.427 |

**Source: data analysis**

### *Latent variable analysis*

Based on the conceptual framework shown in Figure 1, seven-factor models of price perception error were created for each product and service using latent variable analysis (Lavaan). Some latent variables from the collected data were excluded from the analysis because they did not add explanatory power to the model. Thus, the remaining latent variables comprise each factor of the model, owing to their explanatory power. The model fit metrics are presented in Table 11. The GFI and AGFI for all models are greater than 0.95 (Hooper et al., 2008). The RMSEA for all models is less than 0.06, and the SRMR is less than 0.08, suggesting a good fit (Hu and Bentler, 1999). However, the CFI and NFI values are less than the recommended threshold of 0.95 (Hu and Bentler, 1999). Although the p-values for the model chi-squares are less than 0.001, possibly related to sample size (Kline, 2015), the other model fit metrics otherwise indicate that the models have an acceptable fit with room for potential improvement through revising the model specification.

**Table 11.** Latent variable analysis model fit metrics

| Model | $\chi^2$ | p-value | GFI | AGFI | NFI | CFI | RMSEA | SRMR | RFI | IFI | PNFI |
|---|---|---|---|---|---|---|---|---|---|---|---|
| 1 lb Domino granulated sugar | 365.143 | 0.000 | 0.995 | 0.992 | 0.717 | 0.842 | 0.048 | 0.058 | 0.645 | 0.850 | 0.572 |
| 50-inch flatscreen TV | 346.525 | 0.000 | 0.995 | 0.993 | 0.728 | 0.859 | 0.045 | 0.056 | 0.659 | 0.865 | 0.581 |
| City bus ride | 374.321 | 0.000 | 0.995 | 0.992 | 0.709 | 0.833 | 0.049 | 0.058 | 0.636 | 0.841 | 0.566 |
| Couch | 341.823 | 0.000 | 0.995 | 0.993 | 0.736 | 0.866 | 0.044 | 0.055 | 0.669 | 0.872 | 0.588 |
| Dishwasher | 353.443 | 0.000 | 0.995 | 0.992 | 0.723 | 0.852 | 0.046 | 0.056 | 0.653 | 0.859 | 0.577 |
| Dryer | 346.525 | 0.000 | 0.995 | 0.992 | 0.728 | 0.859 | 0.045 | 0.056 | 0.659 | 0.865 | 0.581 |
| Nike Airforce shoe | 344.650 | 0.000 | 0.995 | 0.993 | 0.726 | 0.858 | 0.045 | 0.056 | 0.657 | 0.865 | 0.579 |
| One-night economy hotel stay | 340.827 | 0.000 | 0.995 | 0.993 | 0.731 | 0.863 | 0.044 | 0.056 | 0.663 | 0.869 | 0.583 |
| One-way economy JFK-LAX | 337.814 | 0.000 | 0.995 | 0.993 | 0.730 | 0.864 | 0.044 | 0.055 | 0.662 | 0.871 | 0.583 |
| Round-trip economy JFK-CDJ | 340.426 | 0.000 | 0.995 | 0.993 | 0.731 | 0.863 | 0.044 | 0.055 | 0.663 | 0.870 | 0.583 |
| Zegna boots | 345.970 | 0.000 | 0.998 | 0.996 | 0.725 | 0.857 | 0.045 | 0.056 | 0.655 | 0.863 | 0.579 |

**Source : data analysis**

The coefficients of the structural models for price perception errors for each product and service are displayed in Table 12. Positive loadings indicate the underestimation of prices, and negative loadings indicate overestimation of prices. The decision-making factor had a significant positive loading for dryers and 50-inch televisions (p <.05). The online shopping factor had a significant positive loading for Vans shoes (p<.05). The price sensitivity factor had a significant positive loading for Domino sugar. The demographic factor had a highly significant negative loading for couches (p<0.001) and a significant negative loading for one-night economy hotels and dishwashers (p<0.05). The status brands, quality focus, and financial security factors had no significant loadings.

**Table 12.** Structural model coefficients for price perception error models of products and services

| Model | Status Brands | Decision Making | Online Shopping | Price Sensitivity | Quality Focus | Financial Security | Demographics |
|---|---|---|---|---|---|---|---|
| Dryer | -0.019<br>(0.709) | 0.096<br>(0.038)* | -0.002<br>(0.961) | 0.139<br>(0.213) | 0.015<br>(0.593) | -0.042<br>(0.098) | -1.174<br>(0.088) |
| Economy hotel | -0.019<br>(0.490) | 0.018<br>(0.427) | -0.005<br>(0.820) | 0.044<br>(0.465) | -0.016<br>(0.290) | -0.014<br>(0.332) | -0.673<br>(0.050)* |
| Vans | 0.033<br>(0.188) | -0.039<br>(0.092) | 0.036<br>(0.038)* | 0.100<br>(0.078) | -0.003<br>(0.816) | -0.003<br>(0.810) | -0.047<br>(0.864) |
| Onions | -0.049<br>(0.527) | -0.010<br>(0.865) | 0.089<br>(0.083) | 0.023<br>(0.879) | 0.050<br>(0.186) | 0.009<br>(0.803) | 0.285<br>(0.703) |
| Domino sugar | -0.109<br>(0.165) | -0.046<br>(0.432) | -0.020<br>(0.701) | 0.444<br>(0.018)* | -0.048<br>(0.216) | -0.015<br>(0.663) | 0.691<br>(0.405) |
| JFK-CDG | 0.107<br>(0.203) | 0.119<br>(0.258) | 0.035<br>(0.555) | 0.191<br>(0.224) | 0.027<br>(0.533) | 0.001<br>(0.989) | -0.789<br>(0.409) |
| Zegna boots | 0.001<br>(0.873) | -0.002<br>(0.577) | -0.003<br>(0.476) | 0.009<br>(0.407) | -0.003<br>(0.341) | 0.003<br>(0.322) | -0.033<br>(0.589) |
| Couch | -0.009<br>(0.867) | 0.063<br>(0.202) | -0.008<br>(0.844) | 0.187<br>(0.078) | 0.013<br>(0.652) | 0.023<br>(0.428) | -2.729<br>(0.009)** |
| Dishwasher | 0.014<br>(0.789) | 0.088<br>(0.079) | -0.003<br>(0.939) | 0.085<br>(0.443) | 0.006<br>(0.836) | 0.022<br>(0.398) | -1.770<br>(0.039)* |
| JFK-LAX | 0.017<br>(0.600) | 0.049<br>(0.148) | -0.019<br>(0.429) | 0.021<br>(0.784) | -0.001<br>(0.944) | -0.011<br>(0.478) | 0.066<br>(0.848) |
| City bus | -0.158<br>(0.078) | 0.104<br>(0.168) | -0.076<br>(0.244) | -0.186<br>(0.427) | -0.055<br>(0.249) | -0.043<br>(0.322) | -0.326<br>(0.733) |
| 50-inch TV | -0.027<br>(0.709) | 0.134<br>(0.038)* | -0.003<br>(0.961) | 0.195<br>(0.213) | 0.021<br>(0.593) | -0.059<br>(0.098) | -1.648<br>(0.088) |
| Nike | 0.028<br>(0.266) | -0.003<br>(0.891) | -0.019<br>(0.302) | 0.068<br>(0.261) | -0.023<br>(0.090) | -0.011<br>(0.351) | 0.169<br>(0.530) |

Note: * <.05. ** <.01. *** <.001.

Source: data analysis

The coefficients of the latent variables for each structural factor are listed in Table 13. The name of the structural factor is shown first, followed by the name of the latent variable. Across all models, among the latent variables for the decision-making factor, regret frequency had a highly significant negative loading, followed by money, which had a significantly positive loading. For the Demographics factor, education had a very large positive and significant loading across all models except hotel, where it was highly significant, and income also had a large highly significant loading for the couch model. In the Price Sensitivity factor, the latent variable store brand superior had a significant positive loading across most of the models, with the exception of the city bus model with a highly significant positive loading, and coupon importance had a significant and positive loading only for the sugar model within this factor. Finally, in the factor Status Brands, the latent variables trendy and brand loyalty both had highly significant positive loadings across all models.

**Table 13.** Latent variable coefficients for price perception error models of products and services

| Structural Factor - Latent Variable | Dryer | Hotel | Vans | Onion | Sugar | JFK CDG | Zegna | Couch | Dish-washer | JFK LAX | City bus | 50-inch TV | Nike |
|---|---|---|---|---|---|---|---|---|---|---|---|---|---|
| Decision Making - money wise | 0.224 (0.016) * | 0.252 (0.010) ** | 0.269 (0.009) ** | 0.251 (0.014) * | 0.253 (0.012) * | 0.311 (0.005) ** | 0.238 (0.018) * | 0.256 (0.011) * | 0.260 (0.008) ** | 0.277 (0.005) ** | 0.242 (0.008) ** | 0.224 (0.016) * | 0.245 (0.013) * |
| Decision Making - optimism | 1.000 (NA) | 1.000 (NA) | 1.000 (NA) | 1.000 (NA) | 1.000 (NA) | 1.000 (NA) | 1.000 (NA) | 1.000 (NA) | 1.000 (NA) | 1.000 (NA) | 1.000 (NA) | 1.000 (NA) | 1.000 (NA) |
| Decision Making - regret freq | -0.340 (0.002) ** | -0.379 (0.001) *** | -0.389 (0.001) *** | -0.376 (0.002) ** | -0.376 (0.001) *** | -0.451 (0.001) *** | -0.358 (0.003) ** | -0.381 (0.001) *** | -0.383 (0.001) *** | -0.413 (0.001) *** | -0.374 (0.001) *** | -0.340 (0.002) ** | -0.369 (0.002) ** |
| Demo-graphics - education | 5.178 (0.015) * | 4.651 (0.007) ** | 4.883 (0.015) * | 5.170 (0.018) * | 5.302 (0.019) * | 5.171 (0.016) * | 5.055 (0.014) * | 4.807 (0.005) ** | 5.267 (0.019) * | 5.094 (0.015) * | 5.103 (0.016) * | 5.178 (0.015) * | 5.183 (0.017) * |
| Demo-graphics - gender | 1.000 (NA) | 1.000 (NA) | 1.000 (NA) | 1.000 (NA) | 1.000 (NA) | 1.000 (NA) | 1.000 (NA) | 1.000 (NA) | 1.000 (NA) | 1.000 (NA) | 1.000 (NA) | 1.000 (NA) | 1.000 (NA) |
| Demo-graphics - income | 15.637 (0.058) | 12.004 (0.028) * | 13.664 (0.079) | 15.737 (0.075) | 16.623 (0.078) | 15.650 (0.066) | 14.772 (0.060) | 11.521 (0.010) ** | 16.549 (0.069) | 15.087 (0.060) | 15.146 (0.064) | 15.636 (0.058) | 15.743 (0.068) |
| Demo-graphics - region | -1.849 (0.268) | -1.867 (0.212) | -1.874 (0.236) | -1.841 (0.269) | -1.846 (0.280) | -1.857 (0.266) | -1.872 (0.252) | -1.993 (0.192) | -1.791 (0.289) | -1.866 (0.257) | -1.865 (0.258) | -1.849 (0.268) | -1.848 (0.269) |
| Financial Security – employ-ment security | 1.000 (NA) | 1.000 (NA) | 1.000 (NA) | 1.000 (NA) | 1.000 (NA) | 1.000 (NA) | 1.000 (NA) | 1.000 (NA) | 1.000 (NA) | 1.000 (NA) | 1.000 (NA) | 1.000 (NA) | 1.000 (NA) |
| Online Shopping - onlineshop freq | 1.000 (NA) | 1.000 (NA) | 1.000 (NA) | 1.000 (NA) | 1.000 (NA) | 1.000 (NA) | 1.000 (NA) | 1.000 (NA) | 1.000 (NA) | 1.000 (NA) | 1.000 (NA) | 1.000 (NA) | 1.000 (NA) |
| Price Sensitivity – coupon importnc | 0.838 (0.076) | 0.898 (0.081) | 0.837 (0.058) | 0.846 (0.109) | 0.839 (0.049) * | 0.705 (0.244) | 0.736 (0.142) | 0.437 (0.167) | 0.854 (0.052) | 0.938 (0.108) | 1.046 (0.054) | 0.838 (0.076) | 0.942 (0.082) |
| Price Sensitivity – knockoff mprtnc | 1.000 (NA) | 1.000 (NA) | 1.000 (NA) | 1.000 (NA) | 1.000 (NA) | 1.000 (NA) | 1.000 (NA) | 1.000 (NA) | 1.000 (NA) | 1.000 (NA) | 1.000 (NA) | 1.000 (NA) | 1.000 (NA) |
| Price Sensitivity – store brand sprr | 0.544 (0.024) * | 0.565 (0.019) * | 0.529 (0.016) * | 0.548 (0.031) * | 0.488 (0.012) * | 0.458 (0.144) | 0.507 (0.074) | 0.294 (0.093) | 0.581 (0.020) * | 0.580 (0.018) * | 0.641 (0.008) ** | 0.544 (0.024) * | 0.570 (0.015) * |
| Quality Focus – instore shopfrq | 1.000 (NA) | 1.000 (NA) | 1.000 (NA) | 1.000 (NA) | 1.000 (NA) | 1.000 (NA) | 1.000 (NA) | 1.000 (NA) | 1.000 (NA) | 1.000 (NA) | 1.000 (NA) | 1.000 (NA) | 1.000 (NA) |
| Status Brands – brand loyalty | 0.376 (0.000) *** | 0.375 (0.000) *** | 0.376 (0.000) *** | 0.380 (0.000) *** | 0.382 (0.000) *** | 0.371 (0.000) *** | 0.376 (0.000) *** | 0.378 (0.000) *** | 0.374 (0.000) *** | 0.369 (0.000) *** | 0.355 (0.000) *** | 0.376 (0.000) *** | 0.368 (0.000) *** |
| Status Brands - showoff wealth | 1.000 (NA) | 1.000 (NA) | 1.000 (NA) | 1.000 (NA) | 1.000 (NA) | 1.000 (NA) | 1.000 (NA) | 1.000 (NA) | 1.000 (NA) | 1.000 (NA) | 1.000 (NA) | 1.000 (NA) | 1.000 (NA) |
| Status Brands - trendy | 1.275 (0.000) *** | 1.255 (0.000) *** | 1.389 (0.000) *** | 1.338 (0.000) *** | 1.279 (0.000) *** | 1.272 (0.000) *** | 1.286 (0.000) *** | 1.328 (0.000) *** | 1.276 (0.000) *** | 1.227 (0.000) *** | 1.101 (0.000) *** | 1.275 (0.000) *** | 1.221 (0.000) *** |

**Note:** * <.05. ** <.01. *** <.001. (NA). Not available.

**Source: data analysis**

## 4. Discussion

Consumer attitudes affect price perception and price perception errors. The results from Table 4 suggest that attitudes of respondents appear to seek value and price consciousness characterized by the assignment of high importance to saving money, including using coupons, while also balancing the price-quality tradeoff and online shopping opportunities. However, respondents assigned very low importance to the behaviors of status-seeking, trendiness, conspicuous displays of brand consumption and wealth, and how they choose to allocate their money to saving and spending.

When respondents were asked to guess the prices for a list of selected products and services, some interesting patterns emerged. Table 5 suggests that while the lowest variability in the guessed prices was observed for everyday staples and low-value/low-complexity purchases such as onions, sugar, and local transit, the highest variability in the guessed prices was observed for high-value/high-complexity and intangible purchases such as overseas airfares and household appliances and furnishings. For low-complexity purchases, such as food products, the respondents tended to overestimate prices. For medium-complexity goods that can be categorized as household durable products, respondents tended to underestimate prices. For high-complexity goods and services, including luxury products, respondents tended to greatly underestimate their prices. When the difference between the actual price and the guessed price for each product and service is evaluated in Table 6, the magnitude of these overestimations and underestimations becomes clear. These observations suggest that respondents tend to increasingly underestimate what they believe the price of a product or service is as the complexity and value of the product increases.

There were statistically significant relationships between the mean price errors and the mean Likert scores of the respondents' attitudes and preferences in Table 7, suggesting that there are underlying factors that explain the extent of under- and overestimation of prices for individuals. Under ANOVA and Kruskal-Wallis testing, these statistically significant factors were the importance of brand loyalty, substitute products, knockoff products, how financially well-off the respondent household was growing up, the importance of product quality, the tendency to haggle, and level of income, confirming the observations of Gabor and Grainger (1961), García-Salirrosas et al. (2024), and Zeithaml (1988). In general, the results suggest that respondents who assigned high importance to a factor had greater levels of mean underestimation error than those who assigned low importance, with the exception of respondents who came from relatively impoverished households tending to underestimate more than those from well-off households, confirming the observations of Haymond (2022). The results in Table 7 also reveal some intuitive insights with respect to lower underestimation errors and tendencies to make decisions slowly, being wise with money, and the importance of wearing brand names. The tendency to be risk averse, have low levels of financial security, and assign high importance to coupons or to showing off wealth resulted in higher underestimation errors, confirming the observations of Baysal Kurt and Kara (2024), Dias et al. (2021), Folkes and Wheat (1995), and García-Salirrosas et al. (2024), Zhang (2022), and Raghubir (1998, 2024) on these biases in price perception. These insights suggest that personality traits and personal background act as levers or even biases on how prices are perceived at the individual level. This implies that, despite clever marketing or attractive pricing, the value proposition can be threatened because individual tendencies will confound how the price is framed in their mind. This is especially problematic for marketers if the value proposition is not clear.

Most respondents reported that they relied on some kind of price knowledge heuristic to assist in their evaluation of prices. The results in Table 8 suggest that about 70% of respondents either researched prices first or compared a price to that of a similar product that they wanted. Surprisingly, knowledge of a price from a recent purchase and remembering a price from Walmart or Costco are not popular heuristics. Respondents used the features or benefits of a product the least as a price knowledge heuristic, suggesting that while this information informs us of the value of a product or service, it is probably not sufficient to fully contextualize prices. However, the role of the demographic characteristics of respondents suggests important variations in how they use price knowledge heuristics. Respondents with different levels of education used price knowledge heuristics differently, suggesting that the tendency to research prices first increases with an increasing level of education. Whereas half of the respondents who did not complete high school tended to compare the prices of similar products, about a third of bachelor's degree holders and 40% of master's degree holders researched prices first. Half of the respondents with a post-graduate level of education tended to research prices first. Respondents from different age categories used different price knowledge heuristics. While most respondents aged 18 to 34 tended to compare prices to the price of an item that they wanted, most respondents in age categories from age 35 and older tended to research prices first, suggesting that the latter may have more price knowledge than the former, as posited by Estelami & De Maeye (2004). Gender-based differences in the usage of price knowledge heuristics suggest that although over a third of male and female respondents either preferred to research prices first or compare to the price of a product they wanted, males mainly tended to prefer researching prices more than females, and more than comparing prices to a similar product. Income-based differences in price knowledge heuristics suggest that as the level of income increases from $30,000 or less upward, respondents tend to change their heuristic from comparing the price of similar products to researching prices. Respondents appeared to equally prefer either researching prices first or comparing the price of similar products in the $30,000 to $49,999 per year income category, and then preferring to research prices after this level. These observations support Gabor and Grainger's (1961) claim about the effect of income on price perception. Regionally, there was no clear or consistent pattern of price knowledge heuristic usage, and respondents in each region either tended to prefer researching prices first or mostly comparing the prices of similar products. These observations offer a

signal to marketers to be mindful of the purchasing habits of a given socioeconomic group or subset of a market, insofar as consumers perceive prices.

Respondents were asked what heuristic they used to know if a price was a good deal. The results in Table 10 suggest that patterns of perception errors are linked to different types of purchases and that errors increase as purchase complexity increases. For low-complexity goods such as food products, the mean error is lower among respondents who simply remember the price from their last purchase, and the mean error is higher when they rely upon a Walmart/Costco price heuristic, tending to overestimate. For medium-complexity purchases such as furniture, appliances, and sports footwear, the mean error is lower for Nike and Vans shoe purchases because of reliance on the last price remembered. However, for appliance or furniture purchases, the mean error was generally lower when the respondents relied on a Walmart/Costco price heuristic for couches, resulting in underestimation by relying on the last price remembered for dryers and dishwashers. Curiously, for medium-complexity purchases, the mean error was the highest among those that researched prices first and looked at the features or benefits of the product. In the context of product categories, these observations suggest that while respondents rely on the last remembered price from a prior purchase err less for commoditized products, respondents that use a Walmart/Costco heuristic do worse by overestimating the prices of durable goods. This suggests that price information from a recent purchase provides a more accurate frame for price perception than simply comparing a price to that from Walmart or Costco. These errors reaffirm the observations of Tversky and Kahneman (1974) that people do indeed make errors when using heuristics and do not have perfect price knowledge, as posited by Gabor and Grainger (1961).

The results of latent variable analysis modeling produced a seven-factor perception error model that yielded a good and acceptable fit overall. The loadings on each factor have implications for the overestimation and underestimation of prices for different product categories. The demographic factor had the greatest loadings and was negative (overestimation) for durable and high-value items and negative (underestimation) for staples or low-value items, suggesting that this factor plays a major role in price perception error, given the magnitude of the loadings. The decision-making factor had the next greatest loadings, again mirroring the pattern of overestimation and underestimation of high- and low-value product prices, as with the demographic factor. The price sensitivity factor appears to matter the most for staples and low-value products, but very little for high-value products and intangibles with regard to price perception error. The status-seeking, online shopping, financial security, and quality-focus factors do not appear to have a strong effect on price perception errors, as anticipated and in concurrence with Dias et al. (2022); Fenneman et al. (2022), Zeithaml (1988), and Zhang (2022). When the individual latent variables for each factor were examined, certain patterns emerged to identify the drivers behind the important structural factors in the model. Among the latent variables for the demographic factor, both education and income had significant loadings, confirming Gabor and Grainger’s (1961) observations. The decision-making factor was influenced mainly by how often respondents expressed regret and the extent to which they saved and spent money earned, confirming the observations of Chandrashekaran (2020) and Mukherjee et al. (2021). The price sensitivity factor was influenced by the importance of coupons to respondents and whether they felt that store brands were superior to naming branded products, thus confirming de Wulf et al. ’s observations. (2005), Folkes & Wheat (1995), and Raghubir (1998; 2004). Although not significantly contributing to the model itself, the extent to which respondents follow trends and their loyalty to brands were significant latent variables in the status-seeking factor. Overall, the latent variable analysis modeling suggests that prices are perceived differently for low-value items than for high-value items because of individual differences in price sensitivity driven by levels of income and savings habits, the importance of substitute products, coupons, and brand loyalty. The results suggest that perhaps consumers have expectations for the prices of everyday and durable products that may not be reasonably aligned with reality, given the overestimation of these prices. Furthermore, we infer that consumers do not understand high-value product prices for some reason given their underestimation of prices.

The results of this study indicate that regardless of the preferred price knowledge heuristic, there is a systematic disparity between the price that the consumer has in mind for a purchase versus the actual average price in the market. This disparity presents challenges and has implications for businesses and marketers. First, in markets where products have a corresponding and competing knockoff product, consumers willing to buy counterfeit or inferior knockoff products will have a tendency to incorrectly know the price of the real product. This threatens the consumer value proposition of purchasing and owning the real product versus the knockoff and devalues the importance of product quality. In fact, those who attached higher levels of importance to quality tended to make larger errors, suggesting that quality perception is highly subjective. Marketers should take steps to fully differentiate their offerings so that quality can be viewed as an objective. Second, consumers with higher levels of income and brand loyalty appear to have made better price estimates than those without them. This suggests that brand loyalty should be developed and used as a lever so that consumers can better understand the link between the price and value of a product offering in a meaningful way. Third, since consumers who were indifferent to substitute products and had low levels of importance to knockoff products tended to have smaller price estimate errors, this suggests that consumers do not always refer to the prices of inferior products when using price heuristics. This implies that within a set of similar products that could be bought, the consumer will first gravitate toward the product they are the most loyal to before using the prices of an inferior product in their comparison. Finally, the results suggest that consumers do not understand the pricing of intangibles, such as airfares, shoes, luxury goods, and other everyday goods, to the extent that they greatly underestimate the true price. If consumers underestimate the price, their willingness to pay will not be great. Therefore, overcoming this disconnect can only be accomplished by re-evaluating the value proposition and further granularizing the consumer market segment to understand their objections to pay a premium. Brand loyalty can only be increased if price expectations are correctly aligned with the needs of consumers.

Price knowledge heuristics present specific challenges to marketers and businesses. First, since price knowledge heuristics do not effectively reduce systematic errors in price perception, these insights expose that consumers essentially do not know reliable reference prices or lack knowledge of a particular market and infrequent purchases, as is the case with premium or complex purchases such as luxury products and air fares. To overcome this problem, marketers should seek pricing policies that leverage the comparison of prices from competitors and even decompose the value of a product so that consumers can create a kind of reference or anchor. Second, marketing content should be created to help consumers overcome price sensitivity through value-oriented comparisons and, therefore, develop trust in the value proposition. Third, promotional content should be leveraged to target a specific market segment that is reluctant to make purchases due to price perception errors and motivate purchases through price match guarantees when competitors have set the anchor price. Fourth, messaging to consumers should be differentiated based on the competitiveness of the price for commoditized products and the value for high-price or infrequent purchases. Fifth, leverage consumer behavioral data and feedback from promotional efforts to better target offerings according to the price sensitivity of consumers.

To act strategically upon the above recommendations, marketers and businesses can leverage the demographic nuances of price-perception knowledge to effectively craft different approaches. First, while older consumers tend to research prices or remember the last price they saw, and those in younger age brackets tend to compare prices, marketing messages can use these insights to offer special promotions to these age groups. Aspects of price stability and brand loyalty can be used to appeal to older consumers and deals based on competitive price comparisons to appeal to younger consumers. Second, regional variations in price perception heuristics should motivate marketers to tailor marketing messages to each region that recognizes this particular shopping habit. Third, brand loyalty can be increased if there is increased personalization, wherein marketers somehow incorporate their purchase history so that they can track product purchases and see savings. Finally, since higher levels of education and income are associated with researching prices first, and lower levels of education and income are associated with directly comparing prices, marketing content should be easily understood in both contexts.

## 5. Directions for future research

There is potential for more robust results through new or varied approaches in future research. First, a mixed-methods approach using qualitative data from respondents to give context to qualitative results could offer greater insight into why people perceive prices in the way they do. Accompanying sentiment analysis and machine learning of respondents' textual input could be interpreted in tandem with Likert scores to find groupings of perception methods used by respondents. Second, ongoing work should be conducted to provide a longitudinal profile of how price perception changes over time, particularly regarding how consumers adapt to changes in the retail landscape, economic conditions, and attitudes toward their own lives and their role in the economy. For example, do consumers perceive prices to be better when inflation increases?

## 6. Limitations

The limitations of this study must be acknowledged. First, the sample size used in this study may have an impact on the measurable effects of price perception error to the extent that there is potential for overstatement (Fenneman et al., 2022). Second, since this study used a unique approach to measure the dimensions of respondent attitudes and preferences within the conceptual framework, this may result in the potential of not being able to directly compare results from other authors. Third, because the results presented in this study are based on observations from one moment in time, readers must remember that consumer preferences should not be assumed to remain static or constant throughout time. Therefore, additional studies are required to determine how changing preferences and attitudes alter price perception error over time. Fourth, the conceptual framework focused on key groupings of behavioral dimensions and attitudes of respondents. However, the latent variables for each factor were not exhaustive, and other unmeasured variables could well explain differences in price perception error, such as cultural background, hoarding tendencies, and other thrifty behavioral tendencies. The study was limited to key factors and latent variables in the interest in not creating a cumbersome model or introducing variables with low explanatory power. However, future studies could examine these factors. Finally, despite having taken the appropriate measures to hold respondents accountable by inserting an 'honor code' and explaining that results would be excluded if respondents were found to be cheating or failing attention checks, some respondents were found to fail attention checks or try to skirt the survey controls, such as being found to be not part of the sample population. For this reason, researchers must carefully check and exclude such results, as has been done in this study, and sites such as Amazon Mechanical Turk and Clickworker must be used with caution to ensure data quality.

# APPENDIX A
## Questionnaire

Section 1 of 7

### Attitudes toward pricing

This survey is to ask you about your beliefs and attitudes toward prices of products and services. Please do not rush through the survey and answer honestly. We hope that you find it interesting. Thank you.

Section 2 of 7

Consent

Please read carefully.

*

1. Consent: in order to proceed, you must provide consent. Your responses will be anonymized. No identifying information will be asked, collected, or stored except your age, gender, education, income level, region, and other general characteristics. Your worker ID number will be collected.

To participate, you must be 18 years or older and live in the United States.

Please note: if you answer attention check questions incorrectly, your responses will be flagged.

Do not speed through the survey.

Multiple attempts to take the same survey will be flagged.

Flagged responses will be reported to Clickworker and may result in nonpayment due to work being deficient or unacceptable as per Clickworker Terms and Conditions section 4.3 and 5.3. So, please work honestly.

The survey will ask you to enter information in questions based on your best guess. Do not research or use search tools. This study is about how you perceive prices.

If you do not consent, the survey will end and no data will be collected.

Please make your selection now.

- ◯ Yes, I consent and wish to proceed
- ◯ No, I do not consent and wish to exit.

Section 3 of 7

Agree to terms - honor code.

You must agree to the terms listed below before beginning the survey. If you do not, this survey will terminate and it will exit. If you fail to abide by the terms, Clickworker will be notified and payment will be denied. To ensure that your responses will be payable, please verify your Clickworker ID.

Please enter your Clickworker ID: *

Short-answer text

Please re-enter your Clickworker ID: *

Short-answer text

You agree to answer questions honestly. *

- Yes, I agree to answer questions honestly.
- No, I do not agree.

You agree to not cheat, speed through, or use tools to complete the survey. *

- Yes, I agree to not cheat, speed through or use tools to complete the survey.
- No, I do not agree.

Section 4 of 7

Survey

Thank you for taking this survey. It is hoped that you will find it interesting and fun. Please do not use any aids or look up information. Please use your best guesses. Attention check questions will be asked. Do not speed through the survey.

How important is saving money to you?

Multiple choice

- Definitely unimportant
- Somewhat unimportant
- Neither important nor unimportant
- Somewhat important
- Definitely important
- Add option or Add "Other"

Required

Do you consider yourself wise with money? *

- Definitely unwise
- Somewhat unwise
- Neither wise nor unwise
- Somewhat wise
- Definitely wise

Are you loyal to one or more brands? *

- Definitely disloyal
- Somewhat disloyal
- Neither loyal nor disloyal
- Somewhat loyal
- Definitely loyal

When it comes to the timing of a purchase, how quickly do you like to make decisions about buying something? *

- ◯ Definitely slow to decide
- ◯ Somewhat slow to decide
- ◯ Neither quick nor slow
- ◯ Somewhat quick to decide
- ◯ Definitely quick to decide

When shopping for something, how important are substitute products or services to you? (example: name brands versus off-brands or no-name) *

- ◯ Definitely unimportant
- ◯ Somewhat unimportant
- ◯ Neither important nor unimportant
- ◯ Somewhat important
- ◯ Definitely important

Pick the item that does not belong in this list. *

- ◯ Cheese
- ◯ Milk
- ◯ Curds
- ◯ Concrete
- ◯ Cream

When shopping for something, how important is it for for you that products or services to be available online to buy? *

- ◯ Definitely unimportant
- ◯ Somewhat unimportant
- ◯ Neither important nor unimportant
- ◯ Somewhat important
- ◯ Definitely important

How important are knockoff or imitation products to you when shopping? *

- Definitely unimportant
- Somewhat unimportant
- Neither important nor unimportant
- Somewhat important
- Definitely important

Do you consider yourself to be optimistic about your future? *

- Definitely pessimistic
- Somewhat pessimistic
- Neither optomistic nor pessimistic
- Somewhat optomistic
- Definitely optomistic

How often do you shop online? *

- Definitely seldom
- Somewhat seldom
- Neither often nor seldom
- Somewhat often
- Definitely often

Do you think that store-brand products are better than brand-name products? *

- Store-brand products are definitely worse than brand-name products
- Store-brand products are somewhat worse than brand-name products
- Store-brand products are neither better nor worse than brand-name products
- Store-brand products are somewhat better than brand-name products
- Store-brand products are definitely better than brand-name products

How important are limited-time offers to you when it comes to buying something you want? *

- ◯ Definitely unimportant
- ◯ Somewhat unimportant
- ◯ Neither important nor unimportant
- ◯ Somewhat important
- ◯ Definitely important

How important to you is the country where a product is made? *

- ◯ Definitely unimportant
- ◯ Somewhat unimportant
- ◯ Neither important nor unimportant
- ◯ Somewhat important
- ◯ Definitely important

Growing up, would you say your parents were financially well-off (had money for the family's needs to live comfortably) or impoverished (poor)? *

- Definitely impoverished
- Somewhat impoverished
- Neither financially well-off nor impoverished
- Somewhat well-off
- Definitely well-off

How important is quality when looking at products that are around the same price? *

- Definitely unimportant
- Somewhat unimportant
- Neither important nor unimportant
- Somewhat important
- Definitely important

If it meant saving a lot of money from a big brand price, how likely are you to buy a product that you know is a knock-off? (example: fake Gucci shoes) *

- Definitely unlikely
- Somewhat unlikely
- Neither likely nor unlikely
- Somewhat likely
- Definitely likely

How often do you shop in-person at a store? *

- Definitely seldom
- Somewhat seldom
- Neither often nor seldom
- Somewhat often
- Definitely often

How important is price to you when looking at products that are around the same quality? *

- ◯ Definitely unimportant
- ◯ Somewhat unimportant
- ◯ Neither important nor unimportant
- ◯ Somewhat important
- ◯ Definitely important

Pick the item that does not belong in this list. *

- ◯ Banana
- ◯ Apple
- ◯ Corgi
- ◯ Pineapple
- ◯ Pear

Some people are concerned about achieving a certain status in society (example: become rich, successful, material possessions). How important is social status for you? *

- Definitely unimportant
- Somewhat unimportant
- Neither important nor unimportant
- Somewhat important
- Definitely important

How important to you is wearing brand names or purchasing a brand because you are a fan of it? *

- Definitely unimportant
- Somewhat unimportant
- Neither important nor unimportant
- Somewhat important
- Definitely important

Some people want to show off what they own as symbols of achievement or wealth, often referred to as flexing. How important is this to you? (example: car, watch, clothing) *

- Definitely unimportant
- Somewhat unimportant
- Neither important nor unimportant
- Somewhat important
- Definitely important

When you earn or get extra money, do you save it or spend it? *

- Save all of it
- Save some of it and spend of it
- Spend all of it

How often do you haggle or try to negotiate a deal or price? *

- Definitely seldom
- Somewhat seldom
- Neither often nor seldom
- Somewhat often
- Definitely often

How important to you is being trendy and buying the latest on-trend things? (clothes, tech devices/gadgets) *

- Definitely unimportant
- Somewhat unimportant
- Neither important nor unimportant
- Somewhat important
- Definitely important

Do you consider yourself to be good at decision-making? *

- Definitely bad at decision-making
- Somewhat bad at decision-making
- Neither good nor bad at decision-making
- Somewhat good at decision-making
- Definitely good at decision-making

How important are coupons or discount codes to your purchases? *

- ◯ Definitely unimportant
- ◯ Somewhat unimportant
- ◯ Neither important nor unimportant
- ◯ Somewhat important
- ◯ Definitely important

When you look at a price, how do you know if the price is a good deal to you? *

- ◯ I look at the features or benefits of the product
- ◯ I compare it to what the prices are at Walmart or Costco for the exact same item/brand
- ◯ I research the prices first to get an idea
- ◯ I remember the price from most recent purchase I made for the same item
- ◯ I compare the price to other products that are like the item I want

How willing are you to take risks in decisions that you make (example: purchases, investments, life choices)? *

- ◯ Definitely unwilling to take risks
- ◯ Somewhat unwilling to take risks
- ◯ Neither willing nor unwilling to take risks
- ◯ Somewhat willing to take risks
- ◯ Definitely willing to take risks

How often do you feel regret after buying something? *

- ◯ Definitely seldom
- ◯ Somewhat seldom
- ◯ Neither often nor seldom
- ◯ Somewhat often
- ◯ Definitely often

How often do you splurge or spend extra to treat yourself to something special? *

- Definitely seldom
- Somewhat seldom
- Neither often nor seldom
- Somewhat often
- Definitely often

Do you feel secure about your employment situation? *

- Definitely insecure
- Somewhat insecure
- Neither secure nor insecure
- Somewhat secure
- Definitely secure

How much would you pay for 1 pound of organic sweet onions. Do not research. Enter your value in US dollars and cents using your best idea or guess. *

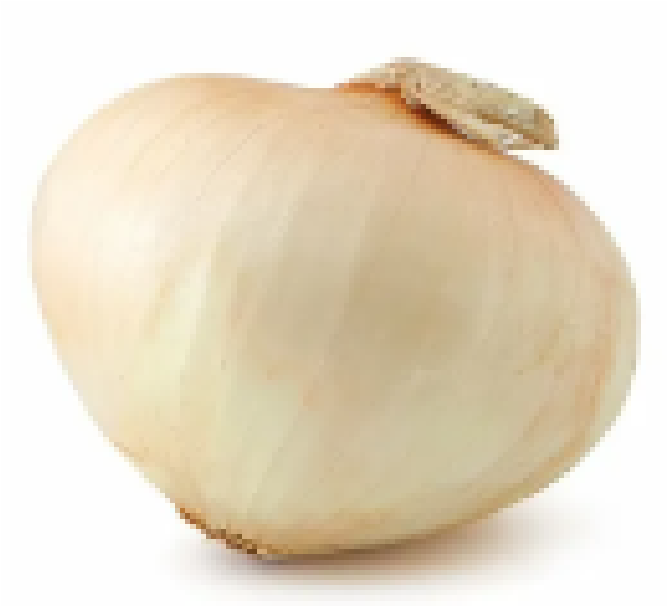

Short-answer text

How much would you pay for this. Do not research. Enter your value in US dollars using your best idea or guess. *

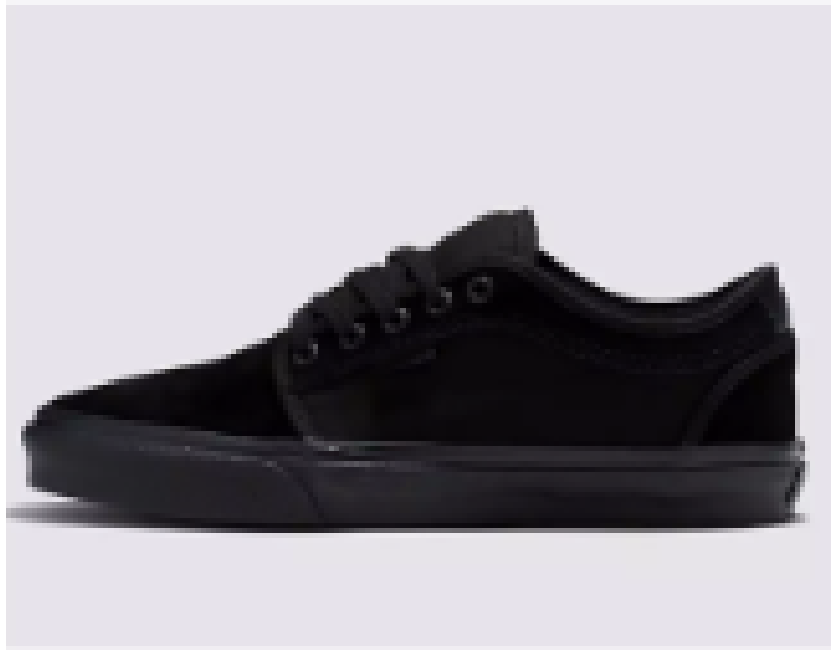

Short-answer text

How much would you pay for this. Do not research. Enter your value in US dollars and cents using your best idea or guess. *

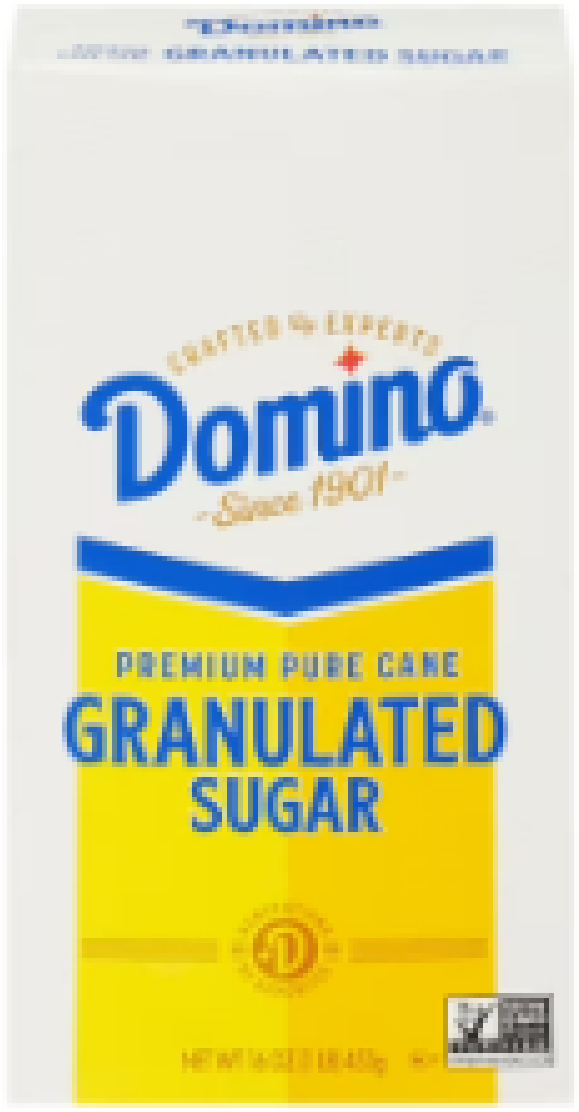

Short-answer text

How much would you pay for this. Do not research. Enter your value in US dollars using your best idea or guess. *

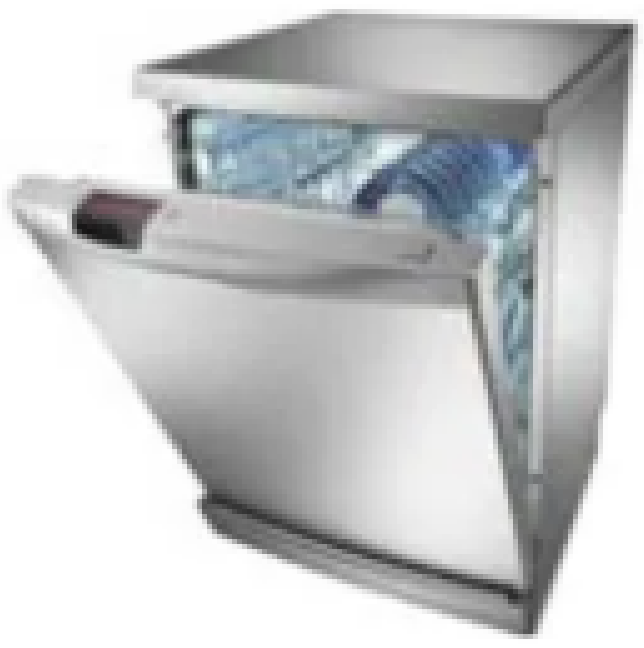

Short-answer text

How much would you pay for this. Do not research. Enter your value in US dollars using your best idea or guess. *

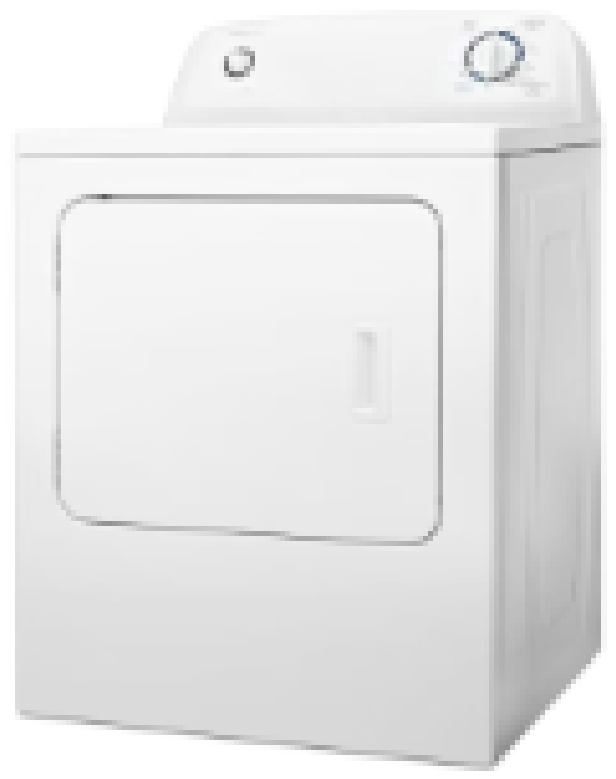

Short-answer text

How much would you pay for this 50-inch flatscreen TV. Do not research. Enter your value in US dollars using your best idea or guess. *

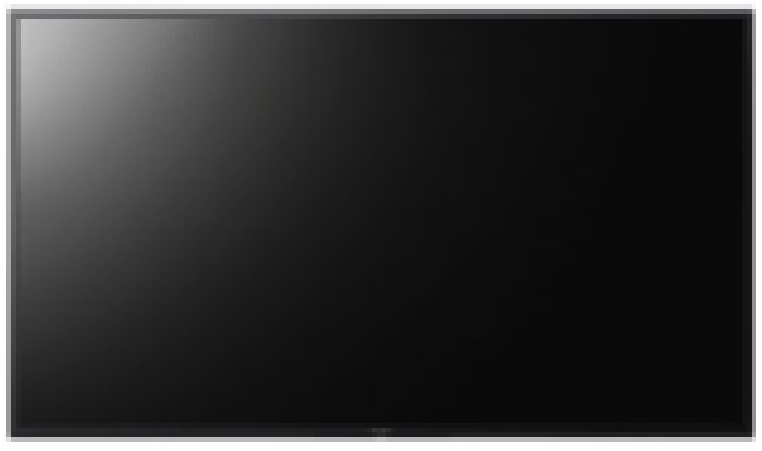

Short-answer text

How much would you pay for this. Do not research. Enter your value in US dollars using your best idea or guess. *

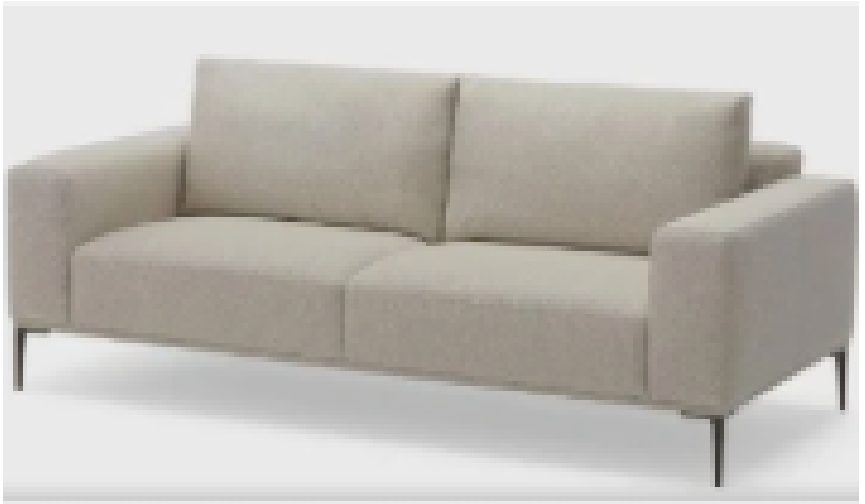

Short-answer text

How much would you pay for this. Do not research. Enter your value in US dollars using your best idea or guess. *

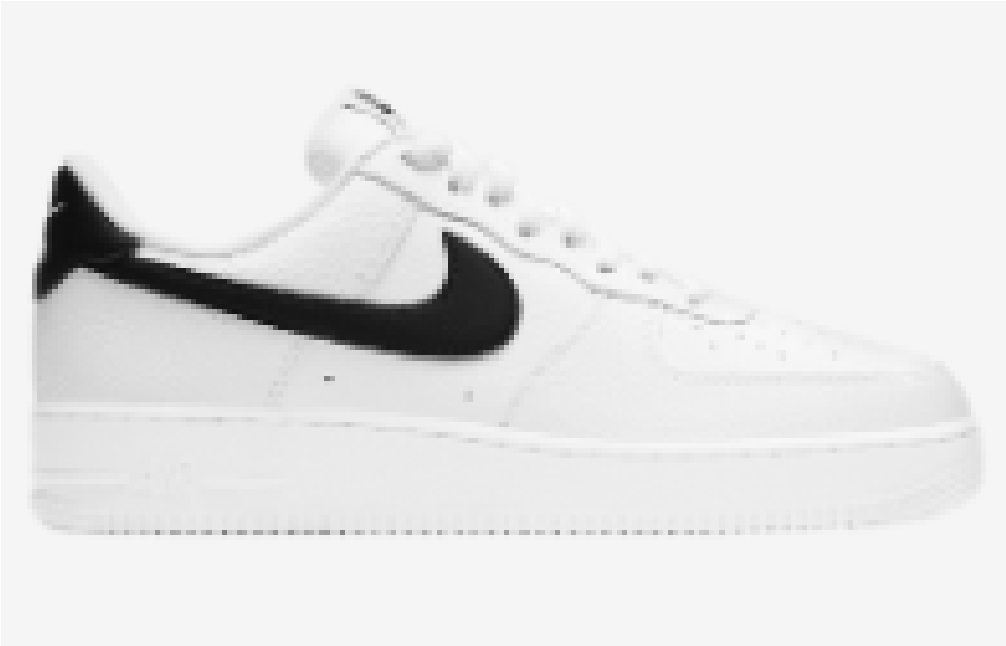

Short-answer text

How much would you pay for this. Do not research. Enter your value in US dollars using your best idea or guess. *

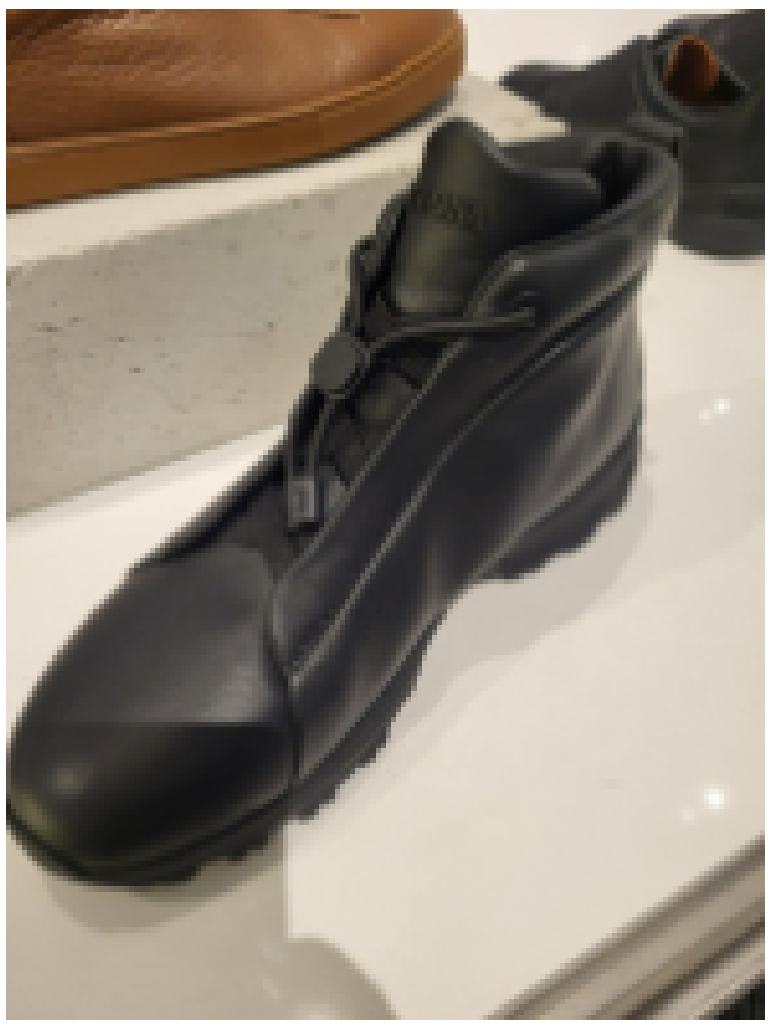

Short-answer text

How much would you pay for a single ride on a city bus. Do not research. Enter your value in US dollars and cents using your best idea or guess. *

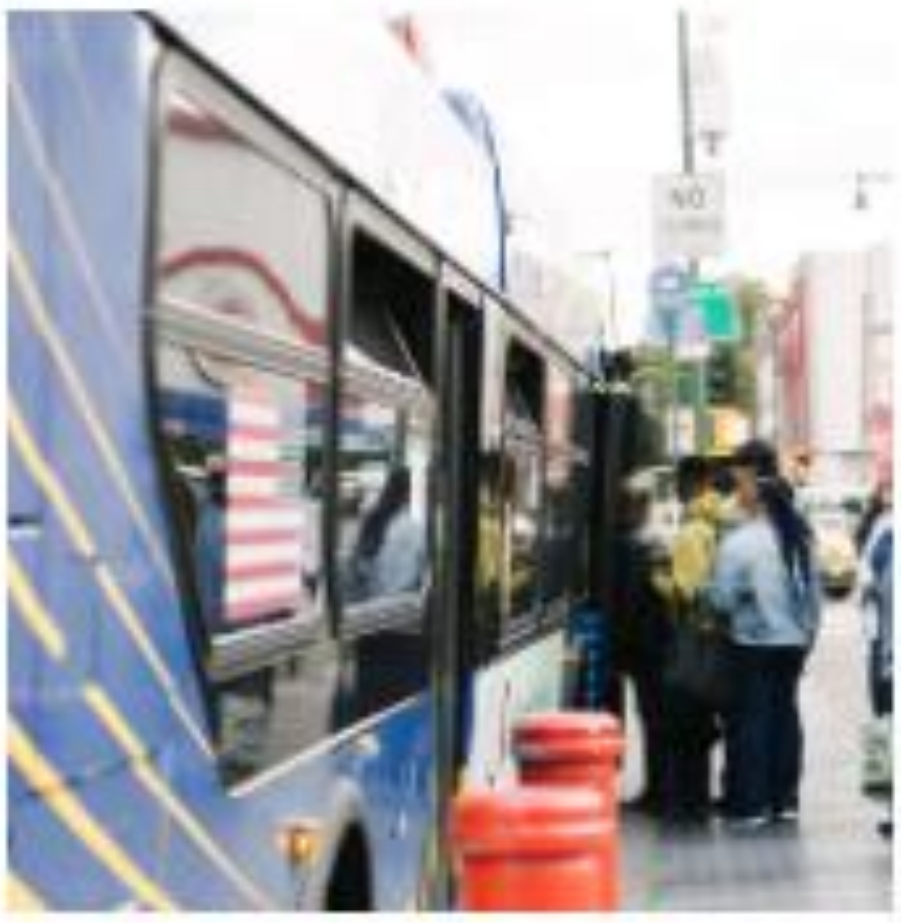

Short-answer text

How much would you for one night in an economy hotel. Do not research. Enter your value in US dollars using your best idea or guess. *

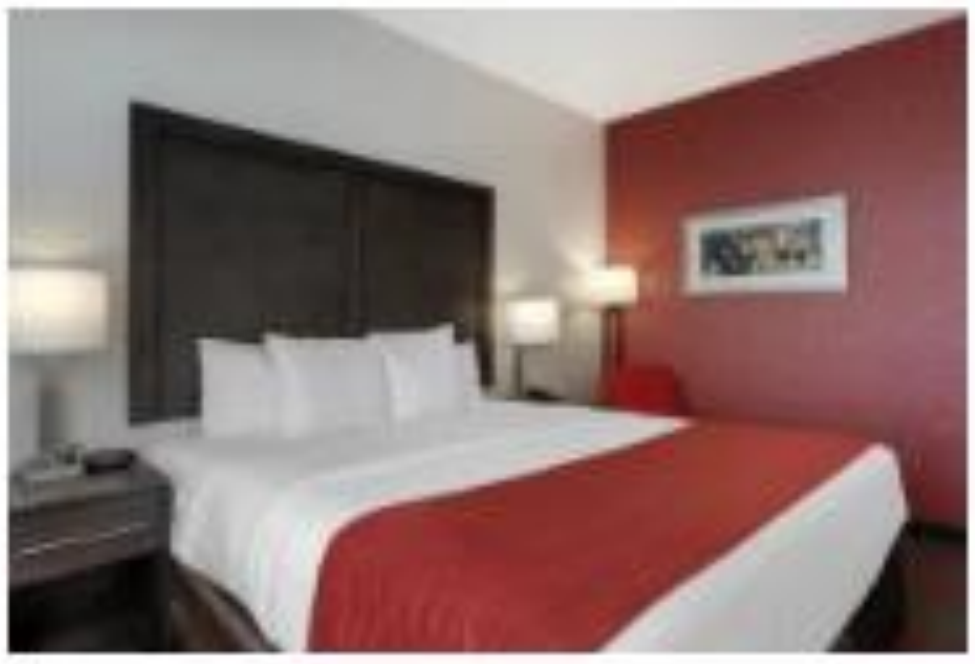

Short-answer text

How much would you for a one-way economy fare from New York to Los Angeles. Do not research. Enter your value in US dollars using your best idea or guess. *

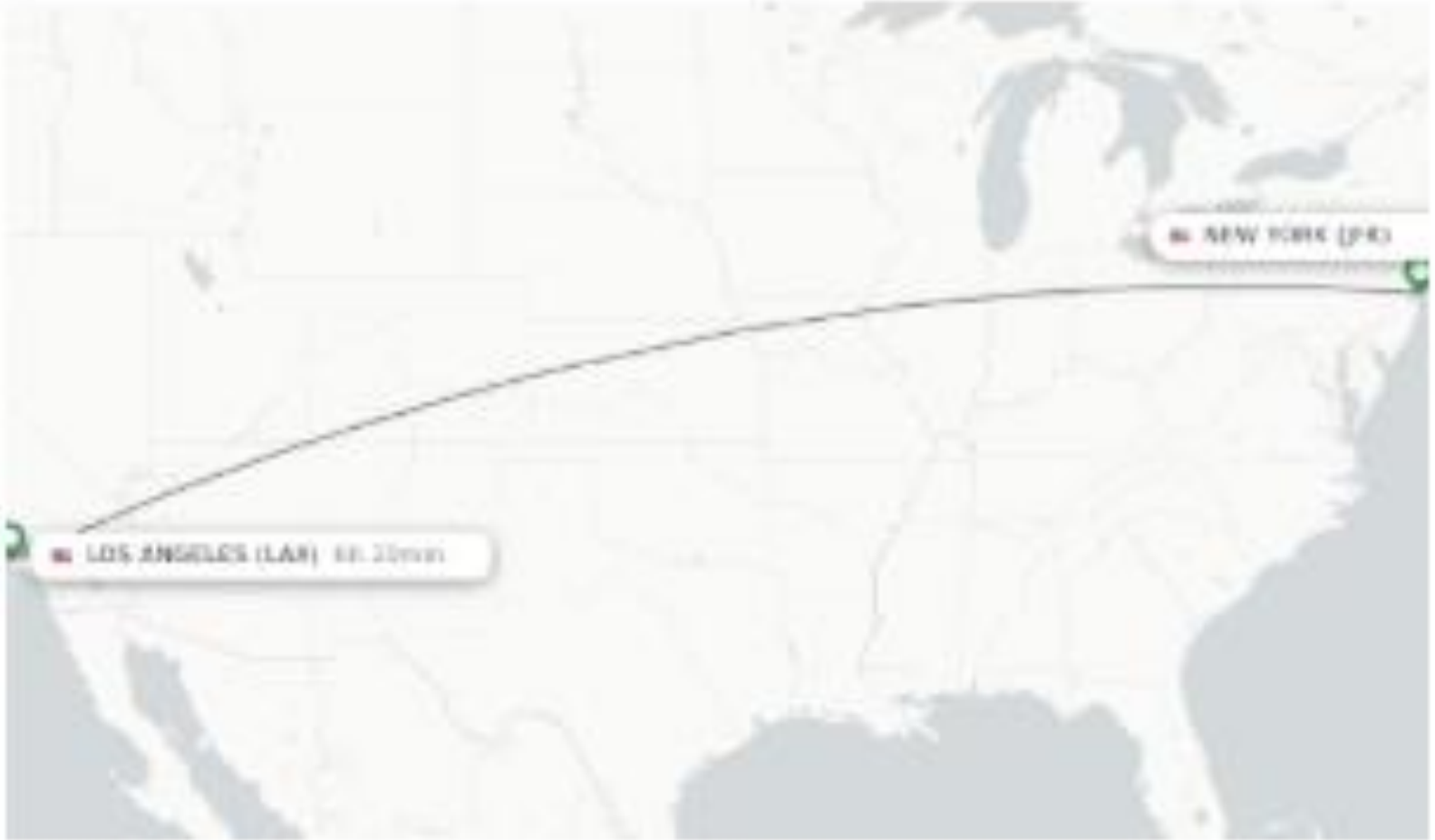

Short-answer text

How much would you for a round-trip economy fare in total for one person from New York to Paris, France, and return. Do not research. Enter your value in US dollars using your best idea or guess. *

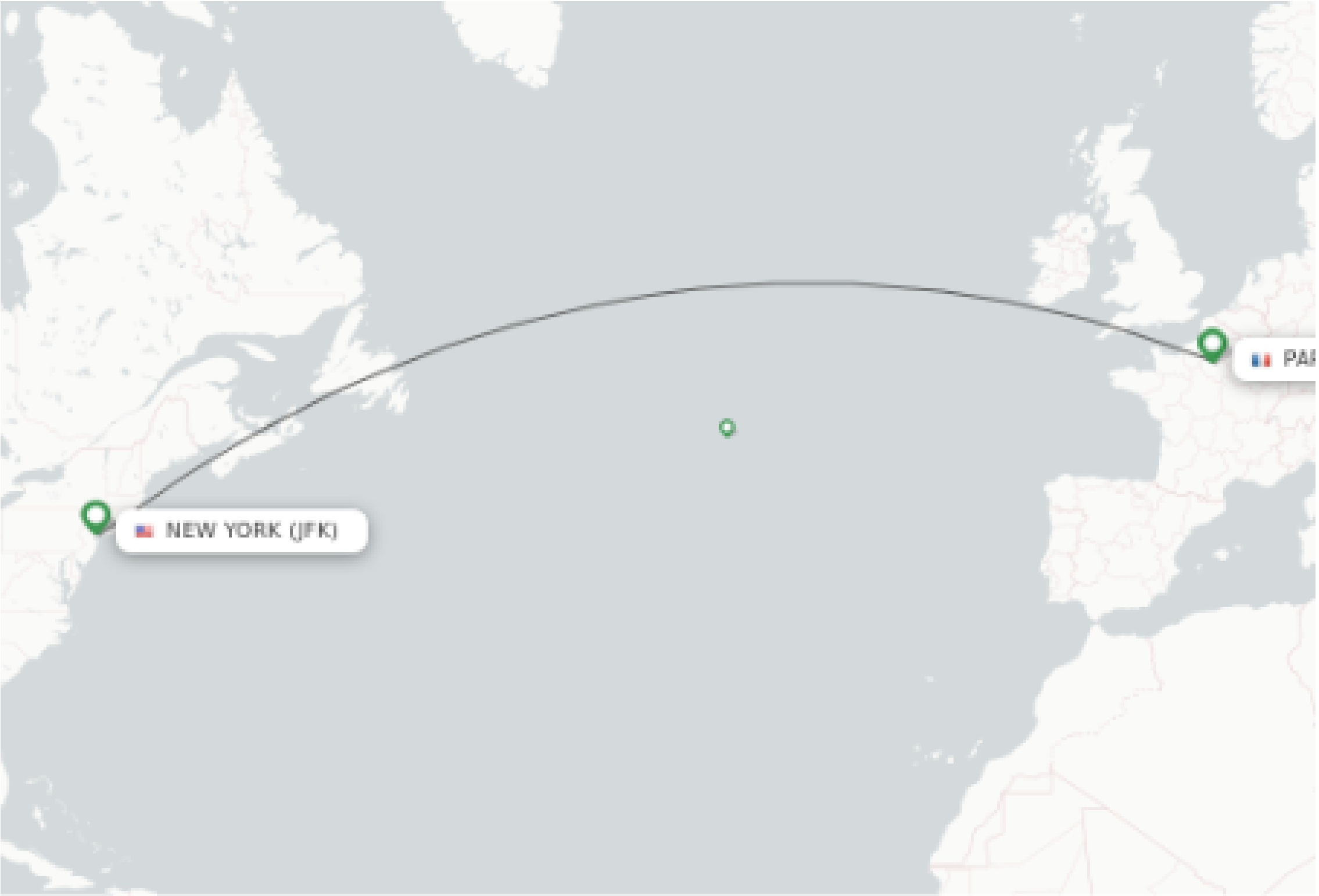

Short-answer text

After section 4 Continue to next section

Section 5 of 7

End of survey - Terminated

Thank you. You did not consent. Your information has not been collected. Survey terminated.

After section 5 Continue to next section

Section 6 of 7

Final questions

Thanks for doing the survey. We're now just going to ask some quick demographic questions about you to finish up.

Please re-verify Clickworker ID? *

This is necessary for payment

Short-answer text

Which age group are you in? *

- 18-24
- 25-34
- 35-44
- 45-54
- 55 and older

What is your gender *

- Male
- Female
- Non-binary

What is your highest level of education? *

- Did not finish high school
- High school graduate
- Some college
- Bachelor's degree
- Master's degree
- Post-graduate degree

What is your level of income? *

- less than $30,000 per year
- $30,0000-$49,999 per year
- $50,000-$69,999 per year
- $70,000 per year or more

What part of the United States do you live in? *

- ◯ Middle Atlantic (NY/NJ/PA)
- ◯ New England: (CT/ME/MA/NH/RI/VT)
- ◯ South Atlantic (DE/DC/FL/GA/MD/NC/SC/VA/WV)
- ◯ East South Central: (AL/KY/MS/TN)
- ◯ West South Central: (AR/LA/OK/TX)
- ◯ Mountain: (AZ/CO/ID/MT/NV/NM/UT/WY)
- ◯ Pacific: (AK/CA/HI/OR/WA)
- ◯ I live outside of the United States

After section 6 Continue to next section

Section 7 of 7

Finished

Many thanks for your participation today!

**READ CAREFULLY - IMPORTANT INSTRUCTION**
**Please copy the following code and paste it into the field provided within your Clickworker task form.**

**Your Clickworker fee cannot be credited without this code!**

Please enter your code: **B6057AD0!y2x?1z** when prompted to ensure payment.

Enter your unique confirmation code. Sharing this code will result in nonpayment for both you and anyone else that you share it with. *

Short-answer text